\documentclass[aps,prl,epsfig,floatfix,showpacs,twocolumn,amssymb,superscriptaddress,amsmath,notitlepage]{revtex4-1}
\usepackage[dvips]{graphicx}
\usepackage{amssymb}
\usepackage{amsmath}
\usepackage{graphicx,color}

\usepackage{float}

\usepackage{enumerate}
\usepackage{mathtools}
\usepackage{stmaryrd}

\usepackage{epstopdf}

\usepackage{textcomp}
\usepackage{gensymb}

\newcommand{\moy}[1]{\left\langle #1 \right\rangle}

\newcommand{\ex}[1]{\mathrm{e}^{#1}}
\newcommand{\sgn}[1]{\mathrm{sgn}(#1)}

\newcommand{\enu}[0]{\ee_{\nu}}

\newcommand{\ee}[0]{\boldsymbol{e}}
\newcommand{\XX}[0]{\boldsymbol{X}}

\newcommand{\rr}[0]{\boldsymbol{r}}
\newcommand{\qq}[0]{\boldsymbol{q}}

\newcommand{\gt}[0]{\widetilde{g}}
\newcommand{\dd}[0]{\mathrm{d}}
\newcommand{\ii}[0]{\mathrm{i}}

\newcommand{\zz}[0]{\mathbf{0}}

\newcommand{\ww}[0]{\boldsymbol{w}}

\newcommand{\beginsupplement}{%
        \setcounter{equation}{0}
        \renewcommand{\theequation}{S\arabic{equation}}%
        \setcounter{figure}{0}
        \renewcommand{\thefigure}{S\arabic{figure}}%
     }

\begin{document}

\title{Nonequilibrium Fluctuations and Enhanced Diffusion of a Driven Particle\\ in a Dense Environment}
\date{\today}

\author{Pierre Illien}
\altaffiliation{Present address: ESPCI Paris, UMR Gulliver, 10 rue Vauquelin, 75005 Paris, France; {pierre.illien@espci.fr}}
\affiliation{Rudolf Peierls Centre for Theoretical Physics, University of Oxford, Oxford OX1 3NP, UK}
\affiliation{Department of Chemistry, The Pennsylvania State University, University Park, PA 16802, USA}

\author{Olivier B\'enichou}
\email{benichou@lptmc.jussieu.fr}
\affiliation{Laboratoire de Physique Th\'eorique de la Mati\`ere Condens\'ee, CNRS UMR 7600, Université\'e Pierre-et-Marie-Curie, 4 Place Jussieu, 75005 Paris, France}

\author{Gleb Oshanin}
\affiliation{Laboratoire de Physique Th\'eorique de la Mati\`ere Condens\'ee, CNRS UMR 7600, Université\'e Pierre-et-Marie-Curie, 4 Place Jussieu, 75005 Paris, France}

\author{Alessandro Sarracino}
\affiliation{Istituto dei Sistemi Complessi-CNR, P.le Aldo Moro 2, 00185, Rome, Italy}
\affiliation{Dipartimento di Fisica, Universit\`a di Roma Sapienza, P.le Aldo Moro 2, 00185, Rome, Italy}

\author{Rapha\"el Voituriez}
\affiliation{Laboratoire de Physique Th\'eorique de la Mati\`ere Condens\'ee, CNRS UMR 7600, Université\'e Pierre-et-Marie-Curie, 4 Place Jussieu, 75005 Paris, France}
\affiliation{Laboratoire Jean Perrin, CNRS UMR 8237, Université\'e Pierre-et-Marie-Curie, 4 Place Jussieu, 75005 Paris, France}

\begin{abstract}
We study the diffusion of a tracer particle driven out-of-equilibrium
by an external force and traveling in a dense environment of arbitrary density. The system
evolves on a discrete lattice and its stochastic dynamics is described
by a master equation. Relying on a decoupling approximation that goes
beyond the naive mean-field treatment of the problem, we calculate
the fluctuations of the position of the tracer around its mean value on a lattice
of arbitrary dimension, and with different boundary conditions. We
reveal intrinsically nonequilibrium effects, such as enhanced
diffusivity of the tracer induced both by the crowding interactions
and the external driving. We finally consider the high-density and
low-density limits of the model and show that our approximation scheme becomes
exact in these limits.
\end{abstract}

\maketitle

\emph{Introduction.---} Biased diffusion in crowded media is
ubiquitous in living systems. At the molecular level, biological
motors are able to overcome thermal fluctuations to achieve directed
motion and perform highly precise functions. At the cellular level,
bacteria are able to self-propel within densely packed biofilms. Both
examples involve a biased, or more generally persistent particle that
moves in a directed manner, and a crowded environment. The description of such systems constitutes a key problem of modern statistical physics \cite{Hofling2013,Chou2011a}. Beyond
fundamental interests, understanding the transport and diffusion
properties of biased particles in complex environments finds
applications in the field of artificial active
matter~\cite{Bechinger2016,Illien2017}, and  in active
microrheology
~\cite{Wilson2011,{Dullens2011},{Puertas2014a}}.  The interplay
between the dynamics of the active agents and their passive
surroundings can trigger self-assembly,
through effective interactions mediated by the quiescent
medium~\cite{Mejia-Monasterio2010, Tanaka2016a}.

Recently, from an analytical perspective, the question of the
diffusion of a biased particle, i.e. the limit case of an active particle with infinite persistence, which interacts with a bath of passive particles has received a growing interest through different approaches \cite{Demery2015,Demery2014}. Here, we focus on the case where the particles interact \emph{via} hardcore interactions and evolve on a lattice. This model is a  variation on exclusion processes, which are  paradigmatic models of nonequilibrium statistical mechanics~\cite{Chou2011a,Mallick2015}.  In the generic situation where the lattice dimension is greater than one and where the density of particles is arbitrary, results are essentially limited to the \emph{mean displacement} of the tracer \cite{Benichou1999a, Benichou2001, Leitmann2013, Benichou2015c}. The \emph{fluctuations} of the tracer position around its mean value received less interest, and results are limited to the case of \emph{fixed} obstacles at low density~\cite{Leitmann2017}, or for  \emph{mobile} obstacles at high density ~\cite{Benichou2013c}. 
Crucially, the fluctuations of the tracer position actually contain information about the environment of the system and its nonequilibrium dynamics, as illustrated by the studies of the diffusion of driven particles in supercooled liquids close to the glass transition \cite{Winter2012,   Schroer2013}, in biased periodic potentials \cite{Reimann2001,Lindenberg2005,Lindner2016}, disordered systems \cite{Reimann2008}, or for active particles \cite{Lindner2008}.
Actually, the problem where
the tracer is not biased
is already highly complex and does not admit an exact solution, although 
an approximate yet very accurate expression
 of the diffusion coefficient as a function of the bath density in 2D was found by Nakazato and Kitahara~\cite{Nakazato1980}.

{
In this Letter, we calculate the fluctuations of the position of the driven tracer around its mean value in the generic case of a bath of arbitrary density and on lattices of dimension $2$ and $3$, which constitute the most physically relevant situations. 
Our analytical approximations are valid both when
the system is infinite in every direction, and when it is confined in
directions perpendicular to the applied bias. Monte-Carlo simulations of the master equation confirm the
accuracy of our closure scheme. 
Remarkably, our approach reveals that the 
diffusion of the tracer can be maximised,  either as a function of the driving force or as a function of  the density of bath particles. 
We emphasize that these effects cannot be predicted 
 within a linear-response description.
We finally show that our approximate expression becomes \emph{exact} in
the high and low density limits, which highlights the consistency and
relevance of our closure scheme.}

\emph{Model.---} We consider the general problem of a biased tracer in
a dynamic environment, i.e. with mobile obstacles of density $\rho$.
The bath particles and
the tracer evolve on a cubic lattice, of spacing $\sigma$ and of
arbitrary dimension $d$, that can be infinite in every direction or
finite with periodic boundary conditions in the directions
perpendicular to the bias. The bath particles perform symmetric random
walks, and jump on adjacent sites with rate $1/(2d\tau^*)$. The tracer
performs a biased random walk, and jumps in direction $\nu$ with rate
$p_\nu/\tau$. We assume hardcore (exclusion) interactions between all
the particles present on the lattice. The set of jump probabilities
$\{p_\nu\}$ is \emph{a priori} arbitrary. However, it can be
convenient to assume that the bias is controlled by an external force
$\boldsymbol{F} = F \ee_1$, and that $p_\nu = \exp(\boldsymbol{F}\cdot
\ee_\nu/2)/Z$, where $Z=\sum_{\mu} \exp(\boldsymbol{F}\cdot
\ee_\mu/2)$ is a normalization constant and $\ee_\mu$ are the base
vectors of the lattice (sums on Greek letter indices run implicitly on $\{\pm 1,\dots, \pm d\}$).

\emph{Analytical approximation.---} The state of the system at a given
time is described by the position of the tracer $\XX$ and the
configuration of the lattice $\eta = \{ \eta_{\rr} \}$, where
$\eta_{\rr}=1$ if site $\rr$ is occupied by a bath particle and 0 otherwise. 
 Enumerating the possible configurations of the system, one
can write the master equation satisfied by the probability
distribution $P(\XX,\eta;t)$ under the form
\begin{equation}
\label{mastereq}
\partial_t P(\XX,\eta ;t)= \mathcal{L}_\text{bath} P  + \mathcal{L}_\text{TP} P ,
\end{equation}
where the terms in the rhs  describe respectively the symmetric diffusion of bath particles and the biased diffusion of the tracer constrained by hardcore interactions. The expression of these operators is given in the Supplemental Material (SM) \cite{SI}. 
The evolution equation for the mean displacement of the tracer can be deduced from the master equation [Eq. \eqref{mastereq}], and was described in previous publications \cite{Benichou2014,Benichou2015c}. We recall it in the SM \cite{SI}. We focus here on the variance of the tracer position in the direction of the bias, defined as 
\begin{equation}
\label{var}
\sigma_X^2(t) \equiv \moy{[X_t-\moy{X_t}]^2} = \moy{{X_t}^2}-\moy{X_t}^2,
\end{equation}
and whose evolution equation is obtained straightforwardly by multiplying Eq. (\ref{mastereq}) by $(\XX \cdot \ee_1)$ and $(\XX\cdot \ee_1)^2$ and summing over all configurations $\XX$ and $\eta$:
\begin{align}
&\frac{\mathrm{d}}{\mathrm{d}t}\sigma_X^2(t)=-\frac{2 \sigma}{\tau}\left[p_1  \widetilde{g}_{\ee_1}(t)-p_{-1} \widetilde{g}_{\ee_{-1}}(t)\right] \nonumber \\
&+\frac{\sigma^2}{\tau}
\left\{ p_1  \left[ 1 -k_{\ee_1}(t) \right]+p_{-1} \left[ 1-k_{\ee_{-1}}(t) \right]\right\},
\label{var}
\end{align}
which holds in dimensions greater than 1, and where we define the density profiles $k_{\rr} \equiv \moy{\eta_{\rr}}$ and the correlation functions $\gt_{\rr} \equiv \moy{(X_t-\moy{X_t}) (\eta_{\rr}-\moy{ \eta_{\rr}})}$  that couple the dynamics of the tracer with that of the bath of particles, and where $\rr$ is evaluated in the frame of reference of the tracer. The diffusion coefficient of the tracer particle, defined as $D \equiv  \frac{1}{2 d } \lim_{t\to\infty} \frac{\mathrm{d}}{\mathrm{d}t}\sigma_X^2(t)$,  can be deduced straightforwardly from Eq. \eqref{var}.

The evolution equations of the density profiles $k_{\rr} $ and of the cross-correlation functions $\gt_{\rr}$ involve higher-order cross-correlation functions, and the infinite hierarchy of equations yielded by the master equation can be closed by the following  mean-field-type decoupling approximations:
\begin{eqnarray}
\moy{\eta_{\rr}\eta_{\rr'}} & \simeq & \moy{\eta_{\rr}} \moy{\eta_{\rr'}} , \label{decoup1}  \\
\moy{\delta X_t \eta_{\rr} \eta_{\rr'}}  & \simeq & \moy{\eta_{\rr}}  \moy{\delta X_t \eta_{\rr'}} + \moy{ \eta_{\rr'}} \moy{\delta X_t \eta_{\rr}}.  \label{decoup2} 
\end{eqnarray}
obtained by writing each random variable $x$ as $x= \moy{x}+\delta x$ and neglecting terms of order $\mathcal{O}(\delta x^2)$ [Eq. \eqref{decoup1}] and $\mathcal{O}(\delta x^3)$ [Eq. \eqref{decoup2}]. 
We  emphasize here that these approximations go beyond naive
mean-field, as the density profiles $\moy{\eta_{\rr}}$ are not
replaced by their spatial average $\rho$.
This closure scheme yields
closed evolution equations for the density profiles and
cross-correlation functions \cite{Illien2015}. Noticing that
$\lim_{|\rr|\to\infty}k_{\rr} = \rho$, i.e. the density profiles relax
to their spatial average far from the tracer, we define the
quantities $h_{\rr} = k_{\rr}-\rho$ and will use the notation
$h_\mu \equiv h_{\ee_\mu}$.

Using discrete Fourier transforms \cite{SI}, we find that, in the stationary limit $t\to \infty$, the density profiles $h_{\rr}$ and the cross-correlation functions $\gt_{\rr}$ obey the equations
\begin{equation}
 \label{systh} 
 \mathcal{A} h_{\rr} = \sum_{\nu} A_\nu h_\nu \nabla_{-\nu} \mathcal{F}_{\rr}-\rho(A_1-A_{-1})(\nabla_1-\nabla_{-1})\mathcal{F}_{\rr},
 \end{equation}
\begin{widetext}
\begin{eqnarray}
&&\gt_{\rr} = \frac{1}{\mathcal{A}}\left \{ \sum_\mu \left( A_\mu-\frac{2d\tau^*}{\tau}p_\mu h_\mu\right)\gt_\mu\nabla_{-\mu} +\frac{2d\tau^*}{\tau} \left[  \rho \sum_{\epsilon = \pm1}\epsilon p_\epsilon \gt_{\epsilon}(\nabla_1-\nabla_{-1})  -\sigma \sum_{\epsilon = \pm1}\epsilon p_\epsilon (1-\rho - h_\epsilon)[\rho(\nabla_{\epsilon}+1)+h_\epsilon]  \right] \right \} \mathcal{F}_{\rr}\nonumber\\
&&-\frac{2d\tau^*}{\tau}\frac{1}{\mathcal{A}^2} \left\{  \sum_{\mu}A_\mu h_\mu \nabla_{-\mu}-\rho(A_1-A_{-1})(\nabla_1-\nabla_{-1}) \right\}  \left\{  \sum_{\mu} p_\mu \gt_\mu \nabla_\mu -\sigma \sum_{\epsilon=\pm1}  \epsilon p_\epsilon (1-\rho-h_\epsilon)\nabla_{\epsilon}  \right\}\mathcal{G}_{\rr},  \label{systgt} 
\end{eqnarray}
\end{widetext}
 where we define the discrete gradient operators $\nabla_\mu f_{\rr} \equiv f_{\rr+\ee_\mu} - f_{\rr}$, the coefficients
\begin{equation}
\label{defAnu}
A_\nu \equiv \frac{1+\frac{2d\tau^*}{\tau}p_\nu(1-\rho-h_\nu)}{\sum_{\mu}[1+\frac{2d\tau^*}{\tau}p_\mu(1-\rho-h_\mu)]},
\end{equation}
and their sum $\mathcal{A} = \sum_\mu A_\mu$. The functions
$\mathcal{F}_{\rr}$ are defined as the limits $\mathcal{F}_{\rr} =
\lim_{\xi \to 1} \widehat{\mathcal{P}}(\rr|\zz;\xi)$ where
$\widehat{\mathcal{P}}(\rr|\zz;\xi)$ is the 
generating function associated with the propagator of a random walk
starting from $\zz$ and arriving at site $\rr$ on a $d$-dimensional
lattice with the following evolution rules: the random walk goes in
direction $-1$ with probability $A_1/\mathcal{A}$, in direction $1$ with
probability $A_{-1}/\mathcal{A}$, and in any other direction with probability
$A_2/\mathcal{A}$. In what follows, we will consider two types of lattices: (i)
$d$-dimensional lattices infinite in every direction, (ii) generalized
capillary-like lattices, infinite in the direction of the applied bias
and finite (of size $L$) with periodic boundary conditions in all the
other directions. The Fourier transform of
$\widehat{\mathcal{P}}(\rr|\zz;\xi)$ is simply given by
$\widetilde{\widehat{\mathcal{P}}}(\qq;\xi)=
[1-\xi\lambda(\qq)]^{-1}$, where $\lambda$ is the structure function
of this random walk \cite{Hughes1995,SI}. We finally define
$\mathcal{G}_{\rr} = \lim_{\xi\to1} \frac{\partial}{\partial
  \lambda}\widehat{\mathcal{P}}(\rr|\zz;\xi)$.
We emphasize the generality of Eqs. (\ref{systh})
and (\ref{systgt}), that hold for different lattice geometries
(infinite or bounded), which only affect the expression of the
generating functions $\widehat{\mathcal{P}}$ \cite{SI}. 

The determination of $D$ requires the knowledge of $h_{\pm1}$ and $\gt_{\pm1}$ [see Eq. \eqref{var}]. Although Eqs. (\ref{systh})
and (\ref{systgt}) cannot
be solved explicitly, $h_{\pm1}$ and $\gt_{\pm1}$ can be determined using a numerical procedure that we sketch here, with further details to be found in the SM \cite{SI}. The first step consists in noticing that Eq. (\ref{systh})
 evaluated for $\rr=\ee_1,\ee_{-1}$ and $\ee_2$ 
yields a closed set of three equations for $h_1$, $h_{-1}$ and $h_2$, where we have used that $h_{\mu} = h_2$ for $\mu=\pm2,...,\pm d$ for symmetry reasons, and the explicit expressions of $A_\nu$ [Eq.\eqref{defAnu}] and $\mathcal{F}_{\rr}$ (Eq. (S24) of SM \cite{SI}). This system is solved numerically for any set of parameters. Next, Eq. (\ref{systgt}) is written for $\rr=\ee_1,\ee_{-1}$ and $\ee_2$, which, now that  $h_1$, $h_{-1}$ and $h_2$ are known, provides a closed set of three equations for $g_1$, $g_{-1}$ and $g_2$. Using the explicit expression of $\mathcal{G}_{\rr}$ (Eq. (S25) of SM \cite{SI}), this set of equations can be solved numerically. Finally, this determines  $h_{\pm1}$ and $\gt_{\pm1}$, and allows us to plot the diffusion coefficient $D$ against the different variables (density, force).

Eqs. (\ref{systh}) and (\ref{systgt}), together with the
evolution equation of the variance $\sigma_X^2(t)$ [Eq. (\ref{var})],
constitute the central result of this Letter. 
Using exact Monte-Carlo
samplings of the master equation, the approximations obtained from our
decoupling scheme are shown to be extremely accurate for a wide range
of parameters. Moreover, we show below that our equations yield
the exact expressions of the fluctuations of the tracer position in
the high- and low-density limits. We also note that, in the absence of bias, our expression reduces to that obtained by Nakazato and Kitahara \cite{Nakazato1980}. From this point of view, our approach constitutes a nonequilibrium extension of that key result. It allows us to unveil typically nonequilibrium effects both with respect to the density and the bias experienced by the tracer.

\begin{figure}
\includegraphics[width=\columnwidth]{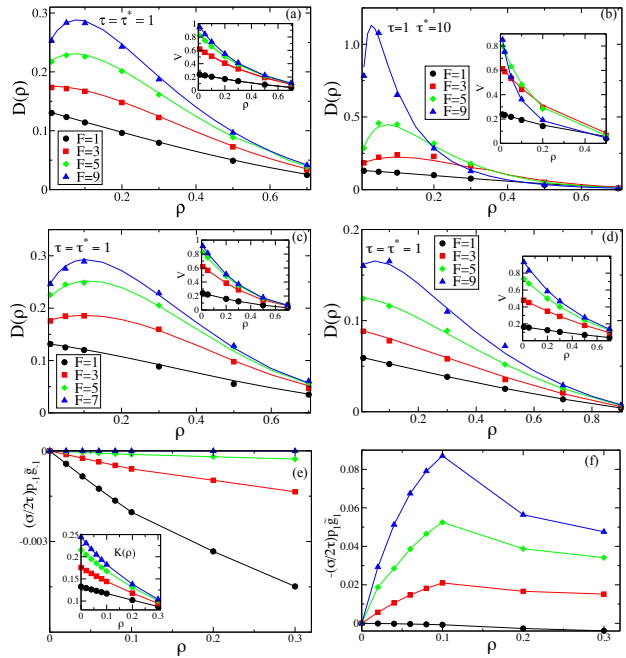}
\caption{(a)-(d): Comparison between  analytical approximations (lines) for
  $D(\rho)$ and numerical simulations (symbols). (a) 2D infinite
  lattice, $\tau=\tau^*=1$.  (b) 2d infinite lattice,
  $\tau=1,\tau^*=10$. (c) Quasi-1d strip-like lattice of width $L=3$,
  with $\tau=\tau^*=1$. (d) 3d infinite lattice, $\tau=\tau^*=1$. The
   approximation is very accurate in a wide range
  of parameters. 
   On each plot, the inset shows the velocity of the tracer particle as a function of the density. {(e) Contributions (analytical approximations) to $D(\rho)$ on a 2D lattice that involve the cross-correlation function $\gt_1$ (e), $\gt_{-1}$ (f) and the density profiles (e, inset)  for the values of $F$ given in panel (a).}}
\label{fig0}
\end{figure}

\emph{Crowding-induced enhanced diffusion.---} Using Eqs. \eqref{systh} and  \eqref{systgt}, we first study the
behavior of $D$ as a function of the particle density $\rho$, at fixed
external force $F$.  As shown in Fig.~\ref{fig0}, which confronts our
analytical approximation with Monte-Carlo simulations of the master
equation, a nonmonotonic behavior is observed for large enough forces.
This means that, counter-intuitively, the diffusivity of the biased
tracer can actually be enhanced by the addition of passive particles
on the lattice.  
To gain insight into this nontrivial behavior, 
 we consider separately the different contributions in the expression of the fluctuations of the tracer position [Eq. (\ref{var})]. While the contribution to the diffusion coefficient involving the density profiles (defined as $K \equiv \frac{\sigma^2}{4\tau}[p_1(1-k_{\ee_1})+p_{-1}(1-k_{\ee_{-1}})])$ and the contribution involving the function $\gt_{\ee_{-1}}$ are systematically monotonous (decreasing) functions of the density [Fig. \ref{fig0}(e)], the contribution involving the cross correlation $\gt_{\ee_{1}}$  becomes non-monotonous for large enough forces [Fig. \ref{fig0}(f)]. This shows that  crowding-induced enhanced diffusion originates from cross-correlations between the position of the tracer and the occupation  of the site located immediately ahead  in the direction of the force, which become more pronounced for an increasing driving force.

\begin{figure}
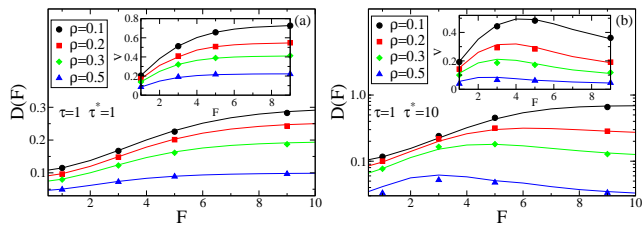

\includegraphics[width=0.48\columnwidth]{fig2a.eps}
\includegraphics[width=0.48\columnwidth]{fig2b.eps}
\caption{Comparison between  analytical approximations (lines) and
  numerical simulations (symbols) for $D(F)$ and $V(F)$ (inset), for $\tau=\tau^*=1$ (a)
  and $\tau=1$ and $\tau^*=10$ (b), for different values of
  $\rho$ in a 2d infinite lattice. Note the nonmonotonic behavior of $D(F)$ for $\tau^*=10$ \cite{note}.}
\label{fig1}
\end{figure}

\emph{Force-induced enhanced diffusion.---} We also study the
dependence of $D$ on the external force, keeping the total
density $\rho$ fixed.  In this case, for large enough values of $\tau^*$ (the typical waiting time of bath particles between two moves), a
non-monotonic behavior of the diffusion coefficient as a function of
$F$ is found (Fig.~\ref{fig1}).
This means that there exists an
optimal value of the external force which produces the maximum of
diffusivity. This kind of behavior is similar to the negative
differential mobility observed in analogous models 
\cite{Leitmann2013,Baerts2013,Benichou2014,Benichou2015c} (see inset of Fig. \ref{fig1}).
Although increasing the driving force reduces the travel time of the tracer between consecutive obstacles, it will increase the times the tracers spends trapped by bath particles if they are slow enough. The tradeoff between these two competing effects results in a non-monotonous dependence of the diffusion coefficient as a function of the driving force, and to the existence of an optimum diffusivity. 
Force-induced enhanced diffusion and negative mobility are found to be related, although the effect is more pronounced for the velocity. For all tested values of parameters, the velocity $V(F)$ and the diffusion coefficient $D(F)$ have the same monotonicity as a function of $F$. Note that, on the contrary,  crowding-induced enhanced diffusion occurs while the velocity is always decreasing with the density.

\emph{High-density limit.---} The high-density limit of the problem
can be studied exactly by relating the statistical properties of the
tracer position to the first-passage densities of the vacancies (empty
sites on the lattice)
\cite{Brummelhuis1988,Brummelhuis1989a,Benichou2002a}. At linear order
in $(1-\rho)$, i.e. when the vacancies have independent dynamics, explicit expressions for the fluctuations of the tracer position have been obtained \cite{Benichou2013c}. In confined systems, this
analysis revealed the existence of a transient regime in which the
fluctuations of the tracer position are superdiffusive, growing as
$t^{3/2}$ on generalized capillaries and as $ t\ln t$ on an infinite
two-dimensional lattice. The tracer ultimately reaches a regular
diffusive regime, after a crossover time that scales as
$1/(1-\rho)^2$, in such a way that the superdiffusive fluctuations can
be long-lived for crowded systems. Importantly, these results can be retrieved using Eqs. \eqref{systh} and  \eqref{systgt}.

First, the transient regime
can be obtained by taking the limit $\rho\to 1$ and then the
long-time limit $t\to \infty$ of the evolution equations for $k_{\rr}$
and $\gt_{\rr}$ \cite{SI}. Using generic relations for propagators on
lattice random walks to simplify the combinations of
$\mathcal{F}_{\rr}$ \cite{Hughes1995,SI}, we obtain the asymptotic
expression for the fluctuations of the tracer, which coincides with
the exact expressions \cite{Benichou2013c}:
\begin{eqnarray}
\label{high_dens_trans}
\sigma_X^2(t) \sim \sigma^2  (1-\rho)\begin{dcases}
\frac{8{a_0}^2}{3 L^{d-1}} \sqrt{\frac{d}{2\pi}} t^{3/2}      & \text{$d$-capillaries}, \\
  \frac{2{a_0}^2}{\pi} t\ln t    & \text{2D lattice},
\end{dcases} 
\end{eqnarray}
where we define
$a_0=\frac{p_1-p_{-1}}{1+\frac{2d\alpha}{2d-\alpha}(p_1+p_{-1})}$. Note
that we considered for simplicity the case where $\tau= \tau^*$, which
corresponds to the discrete vacancy-mediated dynamics described
above. The coefficient $\alpha$ depends on the geometry of the lattice
through the relation $\alpha=\lim_{\xi\to 1
}[\widehat{P}(\zz|\zz;\xi)-\widehat{P}(2\ee_1|\zz;\xi)]$ where
$\widehat{P}(\rr|\rr_0;\xi)$ is the generating function of a symmetric
random walk starting from $\rr_0$ and arriving at $\rr$ on the
considered lattice.

The ultimate diffusive regime is obtained by taking $t\to \infty$ first and ultimately $\rho \to 1$. In the high-density limit, Eqs. (\ref{systh}) and (\ref{systgt}) reduce to linear systems that can be solved explicitly \cite{SI}. We finally obtain 
\begin{eqnarray}
\label{high_dens_ultimate}
\sigma_X^2(t) \sim \begin{dcases}
\frac{2 \sigma^2}{L^{d-1}} \left[  \frac{1}{a_0} + \frac{4 d^2}{L^{d-1}(2d-\alpha)} \right]^{-1}  t  & \text{$d$-capillaries}, \\
  \frac{4\sigma^2{a_0}^2}{\pi}(1-\rho) \ln \left( \frac{1}{1-\rho} \right) t    & \text{2D lattice}.
\end{dcases} \nonumber\\
\end{eqnarray}
The asymptotic expressions of the fluctuations of the tracer position
presented in Eqs. (\ref{high_dens_trans}) and
(\ref{high_dens_ultimate}) then coincide with the results obtained
from the exact approach \cite{Benichou2013c}. This shows that the
decoupling approximation [Eqs. \eqref{decoup1} and \eqref{decoup2}] we propose to treat the master equation of
the problem  is exact in the high-density limit.

\begin{figure}[!t]
\includegraphics[width=0.55\columnwidth,clip=true]{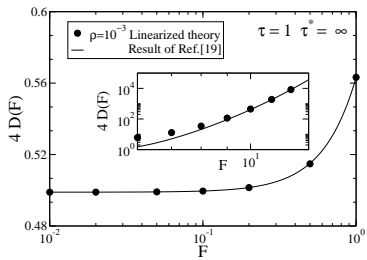}
\caption{Comparison between the result of Ref.~\cite{Leitmann2017} for
  small forces and large forces (inset) and the result from our
  linearized approximation in a 2d infinite lattice with fixed obstacles.}
\label{Franosch_nous}
\end{figure}

\emph{Low-density limit.---} Finally, we consider the low-density
limit of our decoupling approximation. In this limit, the rescaled
density profiles can be expanded as $h_{\nu} = v_{\nu}\rho
+\mathcal{O}(\rho^2)$, where the coefficients $v_{\nu}$ (for
$\nu=\pm1,2$) are the solution of a linear set of three equations \cite{Benichou2014}. By
taking the limit of $\rho\to 0$ of Eq. (\ref{systgt}), we extend this
result to the cross-correlation functions $\gt$ that read $\gt_{\nu} =
u_{\nu}\rho +\mathcal{O}(\rho^2)$, where the coefficients $u_{\nu}$
are the solution of another set of linear equations \cite{SI}.  The
asymptotic expression of the diffusion coefficient $D$ in two
dimensions coincides numerically with the exact analytical solutions
in the limit of fixed obstacles ($\tau^*\to\infty$), that reveal a
non-analytic behavior at small forces and an exponential divergence at
large forces (see Eqs. (16) and (17) from Ref.
\cite{Leitmann2017}). We find an excellent agreement between our
result from the decoupling approximation and the exact expression, as
shown in Fig. \ref{Franosch_nous}. This additional comparison strongly
suggests that our decoupling approximation is exact both in the high-
and low-density limit. We expect this result to hold when the
obstacles can move ($\tau^*<\infty$), as the decoupling approximation
works best when the environment of the tracer is mobile. As a by-product of our approach, we thus obtained an exact expression for the diffusion coefficient of the tracer in the low-density limit.

\emph{Conclusion.---} In this Letter, we studied the statistical
properties of a biased random walker traveling in a passive bath of
particles on a lattice of dimension 2 or more. The master equation of the problem is solved
through a decoupling scheme that goes beyond a naive mean-field
approximation, and we 
calculate the fluctuations of the position of the tracer particle for
an arbitrary set of parameters. We reveal striking counter-intuitive
and intrinsically nonequilibrium effects, namely crowding-induced and
force-induced enhanced diffusion. 
The force-enhanced diffusion is related to the phenomenon of negative
differential mobility \cite{Benichou2014,Benichou2015c}: although increasing the applied force on the tracer can reduce its travel time between different obstacles, it will increase the time it spends trapped by the bath particles it these move sufficiently slowly. The competition between these two effects is at the origin of the non-monotonous behaviour of the diffusion coefficient of the tracer particle.
The effect of density-enhanced diffusion is more subtle and relies on
non-trivial cross-correlations between the tracer and the bath
particles. By studying the different contributions to the diffusion coefficient that are unveiled by our analytical approach, we show that crowding-induced enhanced diffusion originates from the cross-correlations between the tracer position and the occupation of the site ahead, whose contribution becomes dominant when the bias experienced by the tracer is large enough.
We finally show that our decoupling
scheme becomes exact in both the high- and low-density limits, which
validates its relevance.

\emph{Acknowledgments.---} P.I. acknowledges financial support from the U.S. National Science Foundation under MRSEC Grant No. DMR-1420620. The work of O.B. is supported by the European Research Council (Grant No. FPTOpt-277998).

\onecolumngrid

\pagebreak

\beginsupplement

\begin{center}

\textbf{Nonequilibrium Fluctuations and Enhanced Diffusion of a Driven Particle\\ in a Dense Environment}

$\ $

\textit{\textbf{Supplemental Material}}

$\ $

 Pierre Illien,$^{1,2,*}$ Olivier B\'enichou,$^{3,\dagger}$ Gleb Oshanin,$^{3}$ Alessandro Sarracino,$^{4,5}$ and Rapha\"el Voituriez$^{3,6}$

$\ $

$^1$\emph{Rudolf Peierls Centre for Theoretical Physics, University of Oxford, Oxford OX1 3NP, UK}

 $^2$\emph{Department of Chemistry, The Pennsylvania State University, University Park, PA 16802, USA }

$^3$\emph{Laboratoire de Physique Th\'eorique de la Mati\`ere Condens\'ee, CNRS UMR 7600,
Universit\'e Pierre-et-Marie-Curie, 4 Place Jussieu, 75005 Paris, France}

$^4$\emph{Istituto dei Sistemi Complessi-CNR, P.le Aldo Moro 2, 00185, Rome, Italy}

$^5$\emph{Dipartimento di Fisica, Universit\`a di Roma Sapienza, P.le Aldo Moro 2, 00185, Rome, Italy}

$^6$\emph{Laboratoire Jean Perrin, CNRS UMR 8237, Universit\'e Pierre-et-Marie-Curie, 4 Place Jussieu, 75005 Paris, France}

\end{center}

\section{Operators acting on the distribution $P$ in the master equation  [Eq. (1)]}

We denote by $\XX$ the position of the tracer and $\eta_{\rr} \in \{0,1\}$ the occupation number at site $\rr$. $P(\XX,\eta;t)$ is the probability to find the tracer at position $\XX$ with the lattice in configuration $\eta \equiv \{\eta_{\rr}\}$. $\eta^{\rr,\mu}$ is the configuration obtained from $\eta$ by exchanging the occupation numbers of sites $\rr$ and $\rr + \ee_\mu$.

\begin{eqnarray}
2d\tau^*\partial_t P(\XX,\eta;t) =&&
\underbrace{  \sum_{\mu=1}^d\sum_{{\rr}\neq\XX-\ee_\mu,\XX} \left[ P(\XX,\eta^{{\rr},\mu};t)-P(\XX,\eta;t)\right]}_{\equiv \mathcal{L}_\text{bath} P}\nonumber\\
&&\underbrace{+\frac{2d\tau^*}{\tau}\sum_{\mu}p_\mu\left[\left(1-\eta_{\XX} \right)P(\XX-\ee_{\mu},\eta;t) -\left(1-\eta_{\XX+\ee_{\mu}}\right)P(\XX,\eta;t)\right]}_{\equiv \mathcal{L}_\text{TP} P}
\label{eqmaitresse}
\end{eqnarray}

\section{Evolution equation for the mean displacement of the tracer}

Multiplying Eq. \eqref{eqmaitresse} by $(\XX\cdot \ee_1)$ and summing over all configurations $\XX$ and $\eta$, we find
\begin{equation}
\label{ }
\frac{\dd \moy{X_t}}{\dd t} = \frac{\sigma}{\tau} \{ p_1 [1-k_{\ee_1}(t)]-p_{-1} [1-k_{\ee_{-1}}(t)]  \}.
\end{equation}
Noticing that $\lim_{|\rr|\to\infty} k_{\rr} = \rho$, we define the quantities $h_{\rr} = k_{\rr}-\rho$ which are shown to obey Eq. (6) from the main text. Finally, the velocity of the tracer in the stationary limit is defined as $V\equiv \lim_{t\to\infty}  \frac{\dd \moy{X_t}}{\dd t}$.

\section{Derivation of Eqs. (6) and (7)}
\label{SI_eqshgt}

Starting from the master equation and using the decoupling approximations [Eqs. (4) and (5)], we find the equations satisfied by the density profiles $h_{\rr}$:
\begin{itemize}
	\item ${\rr} \notin \{{\bf 0},\pm{\ee_1},\ldots,\pm{\ee_d}\}$
	\begin{equation}
	\label{eq:hbulk}
	2d\tau^*\partial_t h_{\rr} = \widetilde{L} h_{\rr},
	\end{equation}
	
	\item ${\rr} \in \{\pm{\ee_1},\ldots,\pm{\ee_d}\}$
	\begin{equation}
	\label{eq:hBC}
	2d\tau^*\partial_t h_{\enu} = \widetilde{L} h_{\enu} + \rho(A_\nu - A_{-\nu}),
	\end{equation}
\end{itemize}
and the cross-correlation functions $\gt_{\rr}$ \cite{Illien2015}:
\begin{itemize}
	\item for $\rr \notin \{\zz,\ee_{\pm1},\dots,\ee_{\pm d} \}$:
	\begin{align}
	\label{eq:gt_bulk}
	2d\tau^*\partial_t\gt_{\rr}=&\widetilde{L}\gt_{\rr}+\frac{2d\tau^*}{\tau}\sigma\left\{
	p_1(1-\rho-h_1)\nabla_1 h_{\rr}-p_{-1}(1-\rho-h_{-1})\nabla_{-1} h_{\rr}\right\} -\frac{2d\tau^*}{\tau}\sum_\mu p_\mu\gt_{\mu}\nabla_\mu h_{\rr},
	\end{align}
	\item for $\rr=\ee_\nu$ with $\nu \neq \pm 1$:
	\begin{align}
	2d\tau^*\partial_t\gt_{\nu}  = & (\widetilde{L}+A_\nu)\gt_{\nu}+\frac{2d\tau^*}{\tau}\sigma\left\{
	p_1(1-\rho-h_{1})\nabla_1 h_{\nu}-p_{-1}(1-\rho-h_{{-1}})\nabla_{-1} h_{\nu}\right\} \nonumber \\
	&-\frac{2d\tau^*}{\tau}\sum_{\mu}p_\mu\gt_{\ee_\mu}\nabla_\mu h_{\ee_\nu}  -\frac{2d\tau^*}{\tau} \left[p_\nu(\rho+h_\nu)\gt_{\nu}-p_{-\nu} \rho \gt_{\nu}\right],\label{gtBC1}
	\end{align}
	\item for $\rr=\ee_1$ :
	\begin{align}
	2d\tau^*\partial_t\gt_{1}  = & (\widetilde{L}+A_1)\gt_{\nu}+\frac{2d\tau^*}{\tau}\sigma\left\{
	p_1(1-\rho-h_{1})\nabla_1 h_{1}-p_{-1}(1-\rho-h_{{-1}})(\nabla_{-1} h_{1}+\rho)\right\} \nonumber \\
	&-\frac{2d\tau^*}{\tau}\sum_{\mu}p_\mu\gt_{\ee_\mu}\nabla_\mu h_{1}  -\frac{2d\tau^*}{\tau} \left[p_1(\rho+h_1)\gt_{1}-p_{-1} \rho \gt_{-1}\right],\label{gtBC2}
	\end{align}	
	\item for $\rr=\ee_{-1}$ :
	\begin{align}
	2d\tau^*\partial_t\gt_{-1}  = & (\widetilde{L}+A_{-1})\gt_{-1}+\frac{2d\tau^*}{\tau}\sigma\left\{
	p_1(1-\rho-h_{1})(\nabla_1 h_{-1}-\rho)-p_{-1}(1-\rho-h_{{-1}})\nabla_{-1} h_{-1}\right\} \nonumber \\
	&-\frac{2d\tau^*}{\tau}\sum_{\mu}p_\mu\gt_{\ee_\mu}\nabla_\mu h_{-1}  -\frac{2d\tau^*}{\tau} \left[p_{-1}(\rho+h_{-1})\gt_{-1}-p_{1} \rho \gt_{1}\right], \label{bigsyst_last}
	\end{align}	
\end{itemize}
where we defined the operator $\widetilde{L} \equiv\sum_\mu A_\mu\nabla_\mu$ and the quantities $A_\mu\equiv1+\frac{2d\tau^*}{\tau}p_\mu(1-k_{\ee_\mu})$. We will define for simplicity the operators 
\begin{align}
2d\tau^*\partial_t\gt_{\rr} & \equiv \mathcal{L}(\rr),\\
2d\tau^*\partial_t\gt_{\nu} & \equiv \mathcal{L}'(\nu).
\end{align}
Eqs. \eqref{eq:hbulk}-\eqref{bigsyst_last} can be solved by introducing the  auxiliary variable $\ww=(w_1,\dots,w_d)$ and defining the generating functions 
 \begin{eqnarray}
 H(\ww;t) &=& 
 \begin{dcases}
\sum_{r_1=-\infty}^\infty \sum_{r_2,\dots, r_d=0}^{L-1} h_{\rr}(t) \prod_{j=1}^{d} {w_j}^{r_j} & \text{for a generalized capillary}, \\
\sum_{r_1,\dots,r_d=-\infty}^\infty h_{\rr}(t) \prod_{j=1}^{d} {w_j}^{r_j} & \text{for an infinitely extended lattice},
 \end{dcases}   \label{eq:def_generating_H}\\
 G(\ww;t) &=& 
 \begin{dcases}
 \sum_{r_1=-\infty}^\infty \sum_{r_2,\dots, r_d=0}^{L-1}\gt_{\rr} \prod_{j=1}^{d} {w_j}^{r_j} & \text{for a generalized capillary}, \\
 \sum_{r_1,\dots,r_d=-\infty}^\infty \gt_{\rr} \prod_{j=1}^{d} {w_j}^{r_j} & \text{for an infinitely extended lattice}. 
 \end{dcases} 
  \label{eq:def_generating_G}
 \end{eqnarray}
Multiplying Eqs.  \eqref{eq:hbulk} and  \eqref{eq:gt_bulk} by $\prod_{j=1}^d {w_j}^{r_j}$, summing over all lattice sites and using the boundary conditions [Eqs. \eqref{eq:hBC}, \eqref{gtBC1}, \eqref{gtBC2} and \eqref{bigsyst_last}], we find that $H(\ww;t)$ and $G(\ww;t)$ are the solutions of the differential equations
 \begin{eqnarray}
 \label{eq:EDP_H}
 2d\tau^* \partial_t H(\ww;t) &=&  \left[ \frac{A_1}{w_1}+A_{-1}w_1 + A_2\sum_{j=2}^d\left(\frac{1}{w_j} + w_j \right)  - \mathcal{A} \right]H(\ww;t)+K(\ww;t),\\
2d\tau^* \partial_t G(\ww;t) &= & \left[\frac{A_1}{w_1}+{A_{-1}}{w_1}+A_2\sum_{\mu} w_{|\mu|}^{\sgn{\mu}}-\mathcal{A}\right] G(\ww;t) \nonumber \\
& &+ \frac{2d\tau^*}{\tau} \sigma \left[p_1(1-\rho-h_1)\left( \frac{1}{w_1}-1\right)-p_{-1}(1-\rho-h_{-1})(w_1-1)\right]H(\ww;t) \nonumber \\
&&- \frac{2 d \tau^*}{\tau} \left[ \sum_{\mu} p_\mu \gt_\mu \left(\frac{1}{w_{|\mu|}^{\sgn{\mu}}}-1\right) \right]H(\ww;t)-\mathcal{L}_0+\sum_{\mu} w_{|\mu|}^{\sgn{\mu}} \left[\mathcal{L}'(\mu)-\mathcal{L}(\mu)\right], \label{eq:EDP_G}
 \end{eqnarray}
with $\mathcal{A} = A_1+A_{-1} +2(d-1)A_2$ and
 \begin{eqnarray}
 \label{defK}
 K(\ww;t)&\equiv& A_1 (w_1-1)h_{1}+A_{-1} \left(\frac{1}{w_1}-1\right)h_{-1}\nonumber\\
 &+&A_2 \sum_{j=2}^d\left[ (w_j-1)h_j +\left(\frac{1}{w_j}-1\right)h_{-j} \right]+ \rho[A_1 -A_{-1} ] \left( w_1-\frac{1}{w_1}\right),\\
 \mathcal{L}_0&=& \sum_{\mu} A_\mu \gt_\mu + \frac{2d\tau^*}{\tau} \sigma \left[p_1(1-\rho-h_1)-p_{-1}(1-\rho-h_{-1})\right]-\frac{2d\tau^*}{\tau} \sum_{\mu} \gt_\mu h_\mu,
 \end{eqnarray}
and where we used the symmetry relation $A_{\pm 2} =\dots A_{\pm d}=A_2$. In the stationary limit, we find 
\begin{eqnarray}
H(\ww)&=&\frac{K(\ww)}{\mathcal{A}}\frac{1}{1-\lambda(\ww)} ,\label{FourierH}\\
G(\ww)&=&\frac{J_1(\ww)K(\ww)}{\mathcal{A}^2[1-\lambda(\ww)]^2}+\frac{J_0(\ww)}{\mathcal{A}[1-\lambda(\ww)]},\label{FourierG}
\end{eqnarray}
with $\lambda(\ww)=\frac{A_1}{\mathcal{A}} \frac{1}{w_1} +\frac{A_{-1}}{\mathcal{A}} {w_1}  + \frac{A_2}{\mathcal{A}}\sum_{j=2}^d \left( \frac{1}{w_j}+w_j\right)$ and
\begin{align}
J_0(\ww)  \equiv & \sum_{\mu} \left( w_{|\mu|}^{\sgn{\mu}}-1 \right) \left(A_\mu -\frac{2d\tau^*}{\tau}p_\mu h_\mu\right) \gt_\mu -\frac{2d\tau^*}{\tau} \rho \left( w_1-\frac{1}{w_1}\right)(p_1\gt_1-p_{-1}\gt_{-1}) \nonumber\\
&+\frac{2d\tau^*}{\tau} \sigma \left[p_{-1}(1-\rho-h_{-1})\left( \rho{w_1}-h_{-1}\right)-p_{1}(1-\rho-h_{1})(\frac{\rho}{w_1}-h_1)\right], \label{eq:def_J0}\\
J_1(\ww)  \equiv &\frac{2d\tau^*}{\tau} \left\{  \sigma \left[p_1(1-\rho-h_1)\left( \frac{1}{w_1}-1\right)-p_{-1}(1-\rho-h_{-1})(w_1-1)\right] -\sum_{\mu} \gt_\mu \left(\frac{1}{w_{|\mu|}^{\sgn{\mu}}}-1\right)  \right\}. \label{eq:def_J1}
\end{align}

Finally, expressing the generating function variables $w_1,\dots,w_d$ in terms of the Fourier variables:
\begin{itemize}
	\item $w_1=\ex{\ii q_1}$ and $w_j=\ex{\frac{2\ii\pi k_j}{L}}$ ($j\geq2$) for a generalized capillary,
	\item $w_j=\ex{\ii q_j}$ ($0\leq j \leq d$) for an infinitely extended lattice,
\end{itemize}
one can compute the inverse Fourier transform of Eqs. (\ref{FourierH}) and (\ref{FourierG}) in order to retrieve the equations satisfied by $h_{\rr}$ and $\gt_{\rr}$ presented in the main text [Eqs. (6) and (7)].

\section{Expression of the generating functions $\widehat{\mathcal{P}}$}

 \begin{equation}
 \label{eq:def_general_prop_F}
\widehat{\mathcal{P}}(\rr|\zz;\xi)= 
 \begin{dcases}
 \frac{1}{(2\pi)^d}\int_{[-\pi,\pi]^d} \dd q_1\dots \dd q_d \frac{\prod_{j=1}^d \ex{-\ii r_j q_j}}{1-\xi \lambda(q_1,\dots,q_d)}& \text{for an infinitely extended lattice},\\
  \frac{1}{L^{d-1}} \sum_{k_2,\dots,k_d=0}^{L-1} \frac{1}{2\pi}\int_{-\pi}^\pi \dd q \frac{\ex{-\ii r_1 q} \prod_{j=2}^d \ex{-2\ii\pi r_j k_j/L}}{1- \xi \lambda(q,k_2,\dots,k_d)} & \text{for a generalized capillary}, 
 \end{dcases}
 \end{equation}
with the structure factors
\begin{eqnarray}
\lambda(q_1,\dots,q_d) & = & \frac{A_1}{\mathcal{A}} \ex{-\ii q_1 }+ \frac{A_{-1}}{\mathcal{A}} \ex{\ii q_{-1} }+\frac{2A_2}{\mathcal{A}}\sum_{j=2}^d \cos q_j ,\label{lambda_inf}\\
\lambda(q,k_2,\dots,q_d) & = & \frac{A_1}{\mathcal{A}} \ex{-\ii q }+\frac{A_{-1}}{\mathcal{A}} \ex{\ii q }+ \frac{2A_2}{\mathcal{A}}\sum_{j=2}^d \cos \left( \frac{2\pi k_j}{L}  \right) .\label{lambda_cap}
\end{eqnarray}

Using the definition of the functions $\mathcal{F}_{\rr} = \lim_{\xi \to 1} \widehat{\mathcal{P}}(\rr|\zz;\xi)$ and $\mathcal{G}_{\rr} = \lim_{\xi\to1} \frac{\partial}{\partial
  \lambda}\widehat{\mathcal{P}}(\rr|\zz;\xi)$, we find
 \begin{eqnarray}
\mathcal{F}_{\rr}&=& 
 \begin{dcases}
 \frac{1}{(2\pi)^d}\int_{[-\pi,\pi]^d} \dd q_1\dots \dd q_d \frac{\prod_{j=1}^d \ex{-\ii r_j q_j}}{1- \lambda(q_1,\dots,q_d)}& \text{for an infinitely extended lattice},\\
  \frac{1}{L^{d-1}} \sum_{k_2,\dots,k_d=0}^{L-1} \frac{1}{2\pi}\int_{-\pi}^\pi \dd q \frac{\ex{-\ii r_1 q} \prod_{j=2}^d \ex{-2\ii\pi r_j k_j/L}}{1-  \lambda(q,k_2,\dots,k_d)} & \text{for a generalized capillary}, 
 \end{dcases}\\
 \mathcal{G}_{\rr}&=& 
 \begin{dcases}
 \frac{1}{(2\pi)^d}\int_{[-\pi,\pi]^d} \dd q_1\dots \dd q_d \frac{\prod_{j=1}^d \ex{-\ii r_j q_j}}{[1- \lambda(q_1,\dots,q_d)]^2}& \text{for an infinitely extended lattice},\\
  \frac{1}{L^{d-1}} \sum_{k_2,\dots,k_d=0}^{L-1} \frac{1}{2\pi}\int_{-\pi}^\pi \dd q \frac{\ex{-\ii r_1 q} \prod_{j=2}^d \ex{-2\ii\pi r_j k_j/L}}{[1-  \lambda(q,k_2,\dots,k_d)]^2} & \text{for a generalized capillary}, 
 \end{dcases} 
 \end{eqnarray}
where $\lambda$ is given by Eq. \eqref{lambda_inf} for an infinite lattice and by Eq. \eqref{lambda_cap} for a generalized capillary. These expressions of $\mathcal{F}_{\rr}$ and $\mathcal{G}_{\rr}$ are to be used to solve numerically Eqs. (5) and (6) from the main text and to compute the velocity and diffusion coefficient of the tracer particle.

\section{High-density expansion of Eq. (7) }

\subsection{Transient regime -- derivation of Eq. (9)}

We  study the limit where $\rho \to 1$ is taken first, and $t\to \infty$ ultimately, which allows us to calculate the transient regime that precedes the ultimate diffusive regime [Eq. (9)]. To this purpose, we start from the differential equation satisfied by the generating function $G(\ww;t)$ [Eq. \eqref{eq:EDP_G}]. In the limit $\rho \to 1$, at leading order, this equation reduces to
\begin{equation}
2d\tau^* \partial_t G(\ww;t) =  \left[\sum_{\mu} w_{|\mu|}^{\sgn{\mu}}-2d\right] G(\ww;t) +J_0(\ww;t).
\end{equation}
Defining the Laplace transform of any time-dependent function $\psi(t)$ as $\widehat{\psi}(s)= \int_0^\infty \ex{-st}\psi(t) \dd t$, we find 
\begin{equation}
\widehat{G}(\ww;s)  =  \frac{1}{2d(1+\tau^*s)} \frac{\widehat{J}_0(\ww;s)}{1-\frac{1}{1+\tau^*s}\Lambda(\ww)}, \label{FourierLaplaceG}
\end{equation}
where we define 
\begin{equation}
\mathcal{E}_{\boldsymbol{r}} \equiv
\begin{dcases}
 \frac{1}{L^{d-1}} \sum_{k_2,\dots,k_d=0}^{L-1} \frac{1}{2\pi}\int_{-\pi}^\pi \dd q \frac{e^{-\ii r_1q} \prod_{j=2}^d e^{-2\ii\pi r_j k_j/L}}{1- \frac{1}{1+\tau^* s}\Lambda(q,k_2,\dots,k_d)}      & \text{for a generalized capillary}, \\
    \frac{1}{(2\pi)^d}  \int_{[-\pi,\pi]^d} \dd q_1 \dots \dd q_d \frac{ \prod_{j=1}^d e^{-\ii q_i r_i}}{1- \frac{1}{1+\tau^* s}\Lambda(q_1,\dots,q_d)}   & \text{otherwise}.
\end{dcases}
\label{def_E}
\end{equation}
and
\begin{equation}
\Lambda(\ww) \equiv \frac{1}{2 d} \frac{1}{w_1} +\frac{1}{2d} {w_1}  + \frac{1}{2d}\sum_{j=2}^d \left( \frac{1}{w_j}+w_j\right)
\end{equation}
Recalling the definitions of $G$ [Eq. \eqref{eq:def_generating_G}] and $J_0$ [Eq. \eqref{eq:def_J0}], specifying the generating function variable $\ww$ in terms of  Fourier variables (see Section \ref{SI_eqshgt}), and computing the inverse Fourier transform of  Eq. \eqref{FourierLaplaceG}, one gets
\begin{align}
\label{ }
\widehat{\widetilde{g}}_{\rr}(s)=&\sum_{\nu}\widehat{\widetilde{g}}_{\nu}(s) \nabla_{-\nu} \mathcal{E}_{\rr} + \frac{2d\tau^*}{\tau}[p_1\widehat{\widetilde{g}}_{1}(s)-p_{-1}\widehat{\widetilde{g}}_{-1}(s)](\nabla_1-\nabla_{-1})\mathcal{E}_{\rr} \nonumber\\
&-\frac{2d\tau^*}{\tau}\left\{ p_1\left[\frac{1-\rho}{s}-\widehat{h}_1(s)\right](\nabla_1+1)-p_{-1}\left[\frac{1-\rho}{s}-\widehat{h}_{-1}(s)\right](\nabla_{-1}+1)  \right\}\mathcal{E}_{\rr}.
\end{align}
Evaluating this equation for $\rr=\ee_1$, $\ee_{-1}$ and $\ee_2$, we get a system of three equations that can be solved in order to get $\gt_1$, $\gt_{-1}$ and $\gt_2$.  We find that the limit $s\to0$ of the quantities $\mathcal{E}_{\rr}$ are identical to the limit $\xi\to1$ of the propagators $\widehat{P}$. 
Consequently, we find the following sytem:
\begin{eqnarray}
2d \widehat{\widetilde{g}}_1(s) &=& \widehat{\widetilde{g}}_1(s) \nabla_{-1}\widehat{P}_{\ee_1}+\widehat{\widetilde{g}}_{-1}(s) \nabla_{1}\widehat{P}_{\ee_1}+2(d-1)\widehat{\widetilde{g}}_{2}(s)\nabla_2\widehat{P}_{\ee_1} +\frac{2d\tau^*}{\tau}[p_1\widehat{\widetilde{g}}_1(s)-p_{-1}\widehat{\widetilde{g}}_{-1}(s)]\left[\nabla_1\widehat{P}_{\ee_1} - \nabla_{-1}\widehat{P}_{\ee_1}\right] \label{eq:gt_Lap_prov_1} \nonumber\\
&&-\frac{2d\tau^*}{\tau}\sigma\left\{p_1\left[\frac{1-\rho}{s}-\widehat{h}_{1}(s)\right]\widehat{P}_{2\ee_{1}}-p_{-1}\left[\frac{1-\rho}{s}-\widehat{h}_{-1}(s)\right]\widehat{P}_{\zz}\right\},\label{eq:gt_Lap_prov_1}\\
2d \widehat{\widetilde{g}}_{-1}(s) &=& \widehat{\widetilde{g}}_1(s) \nabla_{-1}\widehat{P}_{\ee_{-1}}+\widehat{\widetilde{g}}_{-1}(s) \nabla_{1}\widehat{P}_{\ee_{-1}}+2(d-1)\widehat{\widetilde{g}}_{2}(s)\nabla_2\widehat{P}_{\ee_{-1}}+\frac{2d\tau^*}{\tau}[p_1\widehat{\widetilde{g}}_1(s)-p_{-1}\widehat{\widetilde{g}}_{-1}(s)]\left[\nabla_1\widehat{P}_{\ee_{-1}} - \nabla_{-1}\widehat{P}_{\ee_{-1}}\right]\nonumber\\
&&-\frac{2d\tau^*}{\tau}\sigma\left\{p_1\left[\frac{1-\rho}{s}-\widehat{h}_{1}(s)\right]\widehat{P}_{\zz}-p_{-1}\left[\frac{1-\rho}{s}-\widehat{h}_{-1}(s)\right]\widehat{P}_{2\ee_1}\right\}, \label{eq:gt_Lap_prov_m1}\\ 
2d \widehat{\widetilde{g}}_{2}(s) &=& \widehat{\widetilde{g}}_1(s) \nabla_{-1}\widehat{P}_{\ee_{2}}+\widehat{\widetilde{g}}_{-1}(s) \nabla_{1}\widehat{P}_{\ee_{2}}+\widehat{\widetilde{g}}_{2}(s)\left[\nabla_{-2}\widehat{P}_{\ee_{2}}+\nabla_{2}\widehat{P}_{\ee_{2}}\right]+2(d-2)\widehat{\widetilde{g}}_2(s)\nabla_3\widehat{P}_{\ee_2}\nonumber\\
&&+\frac{2d\tau^*}{\tau}(p_1\widehat{\widetilde{g}}_1(s)-p_{-1}\widehat{\widetilde{g}}_{-1}(s))\left[\nabla_1\widehat{P}_{\ee_{2}} - \nabla_{-1}\widehat{P}_{\ee_{2}}\right]\nonumber\\
&&-\frac{2d\tau^*}{\tau}\sigma\left\{p_1\left[\frac{1-\rho}{s}-\widehat{h}_{1}(s)\right]\widehat{P}_{\ee_{2}+\ee_1}-p_{-1}\left[\frac{1-\rho}{s}-\widehat{h}_{-1}(s)\right]\widehat{P}_{\ee_2+\ee_{-1}}\right\}.\label{eq:gt_Lap_prov_2}
\end{eqnarray}
where $\widehat{P}(\rr|\rr_0;\xi)$ is the generating function associated to the propagator of a symmetric random walk on the considered lattice, starting from $\rr_0$ and arriving at $\rr$. We write for convenience $\nabla_{\nu}\widehat{P}_{\rr} = \lim_{\xi\to1} \nabla_{\nu}\widehat{P}(\rr|\zz;\xi)$. To simplify the system of Eqs. \eqref{eq:gt_Lap_prov_1}-\eqref{eq:gt_Lap_prov_2} a bit further, we also use the following symmetry properties on the quantities $\widehat{P}$ \cite{Hughes1995}:
\begin{eqnarray}
\widehat{P}(\rr|\rr_0;\xi)&=&\widehat{P}(\rr-\rr_0|\zz;\xi), \label{eq:transla_inva}\\
\widehat{P}\left( \sum_{j=1}^d r_j\ee_j |\zz;\xi \right)& =& \widehat{P}\left(-r_i \ee_i+\sum_{\substack{{j=1}\\{j\neq i}}}^d r_j\ee_j|\zz;\xi \right). \label{eq:symmetry}
\end{eqnarray}
The first relation originates from the translational invariance of the lattice, the second one from the symmetry of the random walk described by $\widehat{P}$. We also use the relations
\begin{eqnarray}
\widehat{P}(\zz|\zz,\xi) & =& 1+\frac{1}{d}\left[\widehat{P}(\ee_1|\zz,\xi)+(d-1) \widehat{P}(\ee_2|\zz,\xi)   \right] \label{eq:rel_P_1},\\
\widehat{P}(\ee_1|\zz,\xi) & =& \frac{1}{2d}\left[\widehat{P}(\zz|\zz,\xi)+\widehat{P}(2\ee_1|\zz,\xi)+2(d-1) \widehat{P}(\ee_1+\ee_2|\zz,\xi)   \right], \label{eq:rel_P_2} \\
\widehat{P}(\ee_2|\zz,\xi) & =&  \frac{1}{2d}\left[\widehat{P}(\zz|\zz,\xi)+\widehat{P}(2\ee_2|\zz,\xi)+\widehat{P}(2\ee_1|\zz,\xi)+2(d-2) \widehat{P}(\ee_2+\ee_3|\zz,\xi)   \right],  \label{eq:rel_P_3}
\end{eqnarray}
that are obtained from the generic relation $\widehat{P}(\rr|\rr_0;\xi)=\delta_{\rr,\rr_0}+\frac{\xi}{2d}  \sum_{\mu} \widehat{P}(\rr|\rr_0 +\ee_\mu;\xi)$ \cite{Hughes1995}. 
We introduce the quantities $\alpha=\lim_{\xi\to 1 }[\widehat{P}(\zz|\zz;\xi)-\widehat{P}(2\ee_1|\zz;\xi)]$ and $\beta=\lim_{\xi\to 1 }[\widehat{P}(\zz|\zz;\xi)-\widehat{P}(2\ee_1|\zz;\xi)]$. At leading order when $s \to 0$, one can replace the propagators $\widehat{P}_{\rr}$ by $G_0(1-s)$, where the function $G_0$ is defined as the leading order term of the expansion of the propagators $\widehat{P}(\rr|\rr_0;\xi)$ when $\xi \to 1$:
\begin{equation}
\widehat{P}(\rr|\rr_0;\xi) \underset{\xi\to1}{=}G_0(\xi) +\mathcal{O}(1).
\end{equation}
We deduce that at leading order when $s \to 0$, the system of Eqs. (\ref{eq:gt_Lap_prov_1})-(\ref{eq:gt_Lap_prov_2}) may be written as
\begin{equation}
\mathbf{M}(\delta=0) \begin{pmatrix}
\widehat{\gt}_1(s)   \\
\widehat{\gt}_{-1}(s) \\
\widehat{\gt}_2(s)
\end{pmatrix} = 2d\tau^*  (1-\rho) \frac{p_1-p_{-1}}{1+\frac{2d\alpha}{2d-\alpha} \frac{\tau^*}{\tau}(p_1+p_{-1})} \frac{G_0(1-s)}{s}\begin{pmatrix}
1 \\
1 \\
1
\end{pmatrix},
\end{equation}
where $\mathbf{M}(\delta)$ is defined as 
\begin{equation}
\label{eq:def_Mtilde}
\mathbf{M}=\begin{pmatrix}
-\left[2dp_1\frac{\tau^*}{\tau}(\alpha+2\delta)+2d-\beta-\delta\right]& 2dp_{-1}\frac{\tau^*}{\tau}(\alpha+2\delta)-\alpha+\beta-\delta& \alpha-2\beta   \\
2dp_1\frac{\tau^*}{\tau}(\alpha-2\delta)-\alpha+\beta+\delta& -\left[2dp_{-1}\frac{\tau^*}{\tau}(\alpha-2\delta)+2d-\beta+\delta\right]& \alpha-2\beta   \\
\frac{1}{2(d-1)} \left[ -8\delta p_1 \frac{\tau^*}{\tau} (d-1)-2\beta(d+1) \right.  & \frac{1}{2(d-1)} \left[ -8\delta p_1 \frac{\tau^*}{\tau} (d-1)-2\beta(d+1) \right.  & \frac{1}{d-1} \left[2\beta(d+1)-2d^2-\alpha \right] \\
\hfill +2\delta(d-1)+\alpha-2\beta+2d \Big] & -2\delta(d-1)+\alpha-2\beta+2d \Big] &  
\end{pmatrix},
\end{equation}
with $\delta= d/L^{d-1}$. We finally obtain the expressions of $\widehat{\gt}_1(s)$ and $\widehat{\gt}_{-1}(s)$  in the limit $s\to0$:
\begin{equation}
\widehat{\gt}_{\pm1}(s) \underset{s\to0}{\sim} (1-\rho) \frac{\sigma\tau^*}{\tau} \frac{(p_1-p_{-1})(2d-\alpha)\left(\alpha-2d-\frac{4d\alpha\tau^*}{\tau} p_{\mp1}\right)}{\left[2d-\alpha+\frac{2d\alpha\tau^*}{\tau}(p_1+p_{-1})\right]^2}  \frac{G_0(1-s)}{s}\label{eq:gt_pm1_Lap} 
\end{equation}
The evolution equations for $h_{\rr}(t)$ \cite{Illien2015} can be treated in a similar fashion, to find
\begin{equation}
\label{eq:syst_lin_h_final}
\mathbf{M} \begin{pmatrix}
s \widehat{h}_1(s)   \\
s\widehat{h}_{-1}(s) \\
s\widehat{h}_2(s)
\end{pmatrix} = 2d\frac{\tau^*}{\tau} (1-\rho) (p_1-p_{-1}) \begin{pmatrix}
-\alpha \\
\alpha \\
0
\end{pmatrix},
\end{equation}
and
\begin{equation}
\label{eq:h_1_Lap}
\lim_{\rho \to 1} \frac{\widehat{h}_{1}(s)}{1-\rho} \underset{s\to0}{=} \pm
 \frac{1}{s} \frac{\frac{2d\alpha \tau^*}{\tau}(p_1-p_{-1})}{\frac{2d\alpha \tau^*}{\tau}(p_1+p_{-1})+2d-\alpha}  .
\end{equation}
  Recalling the expression of the second cumulant [Eq. (3)] and considering its Laplace transform, we get
\begin{equation}
\label{eq:Lap_fluc_first_lim}
\mathcal{L}\left[\frac{\mathrm{d}}{\mathrm{d}t}  \sigma^2_X(t) \right](s)  \underset{s\to0}{\sim} 2(1-\rho)\frac{\tau^*\sigma^2}{\tau}\left[\frac{p_1-p_{-1}}{1+\frac{2d\alpha}{2d-\alpha}(p_1+p_{-1})}\right]^2 \frac{G_0(1-s)}{s}.
\end{equation}
Using the following expression of the functions $G_0(\xi)$ \cite{Hughes1995,Brummelhuis1988}
\begin{equation}
\label{ }
G_0(\xi) \underset{\xi\to1}{\sim}
\begin{dcases}
 \frac{\sqrt{d/2}}{L^{d-1}\sqrt{1-\xi}}     & \text{for a capillary of dimension $d$}, \\
     \frac{1}{\pi} \ln \frac{1}{1-\xi} & \text{for a two-dimensional lattice},
\end{dcases}
\end{equation} 
and taking the inverse Laplace transform of Eq. \eqref{eq:Lap_fluc_first_lim}, we finally derive the result presented in the main text [Eq. (9)].

\subsection{Stationary state -- derivation of Eq. (10)}

In the limit where $\rho \to 1$, at leading order in $(1-\rho)$, and in the stationary state we find that Eq. (7) reduces to
\begin{eqnarray}
2d\gt_{\rr} =&& \left\{ \sum_\mu \gt_\mu\nabla_{-\mu} +\frac{2d\tau^*}{\tau}(p_1\gt_{1}-p_{-1}\gt_{-1})(\nabla_1-\nabla_{-1}) \right.\nonumber\\
&&  \left.-   \frac{2d\tau^*}{\tau} \sigma\left\{p_1(1-\rho-h_{1})(\nabla_1+1)-p_{-1}(1-\rho-h_{-1})(\nabla_{-1}+1)]\right\} \right\} \mathcal{F}_{\rr}. \label{gt_highdens_stat}
\end{eqnarray}
In this limit, evaluating Eq. \eqref{gt_highdens_stat} for $\rr=\ee_1,\ee_{-1}$ and $\ee_2$, we find that $\gt_1$, $\gt_{-1}$ and $\gt_2$ are the solutions of the linear system
\begin{eqnarray}
2d \gt_1 &=& \gt_1 \nabla_{-1}\mathcal{F}_{\ee_1}+\gt_{-1} \nabla_{1}\mathcal{F}_{\ee_1}+2(d-1)\gt_{2}\nabla_2\mathcal{F}_{\ee_1}+\frac{2d\tau^*}{\tau}(p_1\gt_1-p_{-1}\gt_{-1})\left[\nabla_1\mathcal{F}_{\ee_1} - \nabla_{-1}\mathcal{F}_{\ee_1}\right] \nonumber\\
&&-\frac{2d\tau^*}{\tau}\sigma[p_1(1-\rho-h_1)\mathcal{F}_{2\ee_{1}}-p_{-1}(1-\rho-h_{-1})\mathcal{F}_{\zz}],\label{eq:system_gt_1}\\
2d \gt_{-1} &=& \gt_1 \nabla_{-1}\mathcal{F}_{\ee_{-1}}+\gt_{-1} \nabla_{1}\mathcal{F}_{\ee_{-1}}+2(d-1)\gt_{2}\nabla_2\mathcal{F}_{\ee_{-1}}+\frac{2d\tau^*}{\tau}(p_1\gt_1-p_{-1}\gt_{-1})\left[\nabla_1\mathcal{F}_{\ee_{-1}} - \nabla_{-1}\mathcal{F}_{\ee_{-1}}\right]\nonumber\\
&&-\frac{2d\tau^*}{\tau}\sigma[p_1(1-\rho-h_1)\mathcal{F}_{\zz}-p_{-1}(1-\rho-h_{-1})\mathcal{F}_{2\ee_{-1}}],\label{eq:system_gt_m1}\\ 
2d \gt_{2} &=& \gt_1 \nabla_{-1}\mathcal{F}_{\ee_{2}}+\gt_{-1} \nabla_{1}\mathcal{F}_{\ee_{2}}+\gt_{2}\left[\nabla_{-2}\mathcal{F}_{\ee_{2}}+\nabla_{2}\mathcal{F}_{\ee_{2}}\right]+2(d-2)\gt_2\nabla_3\mathcal{F}_{\ee_2} \nonumber\\
&&+\frac{2d\tau^*}{\tau}(p_1\gt_1-p_{-1}\gt_{-1})\left[\nabla_1\mathcal{F}_{\ee_{2}} - \nabla_{-1}\mathcal{F}_{\ee_{2}}\right] \nonumber\\
&&-\frac{2d\tau^*}{\tau}\sigma[p_1(1-\rho-h_1)\mathcal{F}_{\ee_2+\ee_1}-p_{-1}(1-\rho-h_{-1})\mathcal{F}_{\ee_2+\ee_{-1}}].\label{eq:system_gt_2}
\end{eqnarray}
In what follows, we study the $\rho \to 1$ limit of these equations in the two situations where the lattice is a generalized capillary, and where it is a two-dimensional lattice.

\subsubsection{Generalized capillaries}

We will use the relation
\begin{equation}
\label{eq:limit_gradients_F}
\lim_{\rho_0 \to 0} \nabla_\nu \mathcal{F}_{\rr} = \lim_{\xi \to 1} \left[ \widehat{P}(\rr+\ee_\nu|\zz;\xi) - \widehat{P}(\rr|\zz;\xi)   \right]+\delta_\nu.
\end{equation}
with
\begin{equation}
\label{ }
\delta_\nu = \begin{cases}
-\frac{d}{L^{d-1}}      & \text{if $\nu=1$}, \\
\frac{d}{L^{d-1}}      & \text{if $\nu=-1$}, \\
0& \text{otherwise}.
\end{cases}
\end{equation}
Contrary to the equation verified by the quantities $(h_1,h_{-1},h_2)$ [Eq. (6)] which only involves \emph{differences} of the functions $\mathcal{F}_{\rr}$, the system (\ref{eq:system_gt_1})-(\ref{eq:system_gt_2}) also involves functions $\mathcal{F}_{\rr}$ alone, which diverge when $\rho \to 1$. In the case of generalized capillaries, we find that 
\begin{equation}
\label{eq:limit_F_seul}
\mathcal{F}_{\rr} \underset{\rho\to 1}{\sim}\frac{1}{L^{d-1} V} \equiv \mathcal{G}(V),
\end{equation}
 where $V$ is the velocity of the tracer \cite{Illien2014,Benichou2014}:
\begin{equation}
\label{eq:vitesse_def_bande_V}
V \underset{\rho \to 1}{\sim} \frac{\sigma}{\tau} \left[ p_1(\rho-1-h_1)-p_{-1}(\rho-1-h_{-1})  \right],
\end{equation}
and vanishes when $\rho \to 1$. Using Eqs. (\ref{eq:limit_F_seul}) and (\ref{eq:limit_gradients_F}), we simplify  Eqs. (\ref{eq:system_gt_1})-(\ref{eq:system_gt_2}).  
With the usual symmetry properties on the quantities $\widehat{P}$ [Eqs. (\ref{eq:transla_inva})-(\ref{eq:symmetry})] as well as the relations (\ref{eq:rel_P_1})-(\ref{eq:rel_P_3}), we rewrite Eqs. (\ref{eq:system_gt_1})-(\ref{eq:system_gt_2}) in terms of the propagators $\widehat{P}(\zz|\zz;\xi)$,  $\widehat{P}(2\ee_1|\zz;\xi)$ and $\widehat{P}(\ee_1|\zz;\xi)$ only. We finally show that $\gt_1$, $\gt_{-1}$ and $\gt_2$ are the solutions of the linear system
\begin{equation}
\label{eq:syst_lin_gt_final}
\mathbf{M}\begin{pmatrix}
\gt_1   \\
\gt_{-1} \\
\gt_2
\end{pmatrix} = 2d\tau^* V (1-\rho) (p_1-p_{-1}) \mathcal{G}(V)\begin{pmatrix}
1 \\
1 \\
1
\end{pmatrix},
\end{equation}
where $\mathbf{M}$ is defined as previously [Eq. \eqref{eq:def_Mtilde}].
At leading order in $\rho\to 1$, this system has the following solutions for $\gt_{\pm1}$:
\begin{equation}
\gt_{\pm 1} \underset{\rho \to 1}{\sim}  \frac{\sigma \tau^*}{L^{d-1}} \frac{\alpha-2d-\frac{4d\alpha \tau^*}{\tau}p_{\mp1}}{\left[2d-\alpha+\frac{2d\alpha \tau^*}{\tau}(p_1+p_{-1})+\frac{4d^2}{L^{d-1}}\frac{\tau^*}{\tau}(p_1-p_{-1})\right]^2}.
\end{equation}
With a similar treatment of Eq. (6), we find that $h_1$, $h_{-1}$ and $h_2$ are the solution of the linear set of equations
\begin{equation}
\mathbf{M}\begin{pmatrix}
h_1   \\
h_{-1} \\
h_2
\end{pmatrix} = 2d \frac{\tau^*}{\tau} (1-\rho) (p_1-p_{-1})\begin{pmatrix}
-\alpha-2\delta \\
\alpha-2\delta \\
-2\delta
\end{pmatrix}.
\end{equation}
We deduce the expressions of $h_{\pm 1}$:
\begin{equation}
\label{ }
h_{\pm1} = \pm (1-\rho) \frac{\frac{2d\tau^*}{\tau}(\alpha+\frac{2d}{L^{d-1}})(p_1-p_{-1})}{\frac{2d\tau^*}{\tau}\alpha(p_1+p_{-1})+\frac{4d^2}{L^{d-1}}\frac{\tau^*}{\tau}(p_1-p_{-1})+   2d-\alpha} 
\end{equation}
Using Eq. (3), this yields the result presented in the main text [Eq. (10)].

\subsubsection{Two-dimensional lattice}

Extending the relation (\ref{eq:limit_gradients_F}) to an infinite two-dimensional lattice by taking $d=2$ and $L\to \infty$, we get:
\begin{equation}
\lim_{\rho_0 \to 0} \nabla_\nu \mathcal{F}_{\rr} = \lim_{\xi \to 1} \left[ \widehat{P}(\rr+\ee_\nu|\zz;\xi) - \widehat{P}(\rr|\zz;\xi)   \right].
\end{equation}
We can also show that, for the case of a two dimensional lattice, 
\begin{equation}
\mathcal{F}_{\rr} \underset{\rho_0 \to 0}{\sim}\frac{2}{\pi} \ln \frac{1}{V},
\end{equation}
where $V$ is the velocity of the TP [Eq. (\ref{eq:vitesse_def_bande_V})]. Using again the symmetry relations presented above [Eqs. \eqref{eq:transla_inva}-\eqref{eq:rel_P_3}], we find that $\gt_1$, $\gt_{-1}$ and $\gt_2$ are the solutions of the system (\ref{eq:syst_lin_gt_final}), where we take $\mathcal{G}(V)=\frac{2}{\pi} \ln \frac{1}{V}$.  We obtain the following solutions of the system at leading order in $\rho \to 1$:
\begin{equation}
\gt_{\pm 1} \underset{\rho \to 1}{\sim} (1-\rho) \frac{\tau^*}{\tau}\sigma \frac{(\alpha-4)(p_1-p_{-1})\left(8\alpha\frac{\tau^*}{\tau}+4-\alpha  \right)}{4-\alpha+4\alpha\frac{\tau^*}{\tau}(p_1+p_{-1})} \frac{2}{\pi} \ln \frac{1}{V} 
\end{equation}
and deduce the expression of the diffusion coefficient presented in the main text in Eq. (10).

\section{Low-density expansion}

We start from the expression of the diffusion coefficient in terms of the density profiles $k_{\rr}$ and the cross-correlation functions $\widetilde{g}_{\rr}$ [Eq. (3)].
In the $\rho \to 0$ limit, we define the functions $v_{\rr}$ and $u_{\rr}$ as follows:
\begin{equation}
k_{\rr}  \underset{\rho\to0}{=}  (1+v_{\rr})\rho + \mathcal{O}(\rho^2) \qquad\qquad ; \qquad\qquad
\gt_{\rr}  \underset{\rho\to0}{=}  u_{\rr}\rho + \mathcal{O}(\rho^2) ,
\end{equation}
so that equation diffusion coefficient reads
\begin{equation}
\label{K_small_rho_exp}
D \underset{\rho\to0}{=} \frac{\sigma^2}{2 d \tau} (p_1+p_{-1}) - \frac{\sigma}{2 d \tau} \rho \{ p_1 [\sigma(1+v_1)+2 u_1]+p_{-1}[\sigma(1+v_{-1})-2 u_{-1}]  \}+ \mathcal{O}(\rho^2) .
\end{equation}
The functions $v_{\nu}$ have been studied before in the low-density limit \cite{Benichou2014}, and are the solutions of the system:
\begin{equation}
\label{ }
\widetilde{D} \widetilde{v}  = (p_1-p_{-1}) \widetilde{\mathcal{F}},
\end{equation}
with
\begin{equation}
\label{ }
\widetilde{v}= \begin{pmatrix}
v_1   \\
v_{-1} \\
v_2
\end{pmatrix}
\qquad\qquad;\qquad\qquad
\widetilde{\mathcal{F}}= \begin{pmatrix}
(\nabla_1-\nabla_{-1}) \mathcal{F}_{\ee_1}   \\
(\nabla_1-\nabla_{-1}) \mathcal{F}_{\ee_{-1}} \\
(\nabla_1-\nabla_{-1}) \mathcal{F}_{\ee_2} 
\end{pmatrix},
\end{equation}
and
\begin{equation}
\widetilde{D}=\frac{1}{2d \left( 1+\frac{\tau^*}{\tau}   \right)} \begin{pmatrix}
\left( 1 + \frac{2d\tau^*}{\tau} p_{1}\right) \nabla_{-1} \mathcal{F}_{\ee_1} -1& \left( 1 + \frac{2d\tau^*}{\tau} p_{-1}\right)\nabla_{1} \mathcal{F}_{\ee_1} & 2  \left( 1 + \frac{2d\tau^*}{\tau} p_{2}\right)\nabla_{2} \mathcal{F}_{\ee_1} \\
\left( 1 + \frac{2d\tau^*}{\tau} p_{1}\right) \nabla_{-1} \mathcal{F}_{\ee_{-1}}  & \left( 1 + \frac{2d\tau^*}{\tau} p_{-1}\right)\nabla_{1} \mathcal{F}_{\ee_{-1}}-1 & 2 \left( 1 + \frac{2d\tau^*}{\tau} p_{2}\right) \nabla_{2} \mathcal{F}_{\ee_{-1}} \\
\left( 1 + \frac{2d\tau^*}{\tau} p_{1}\right)\nabla_{-1} \mathcal{F}_{\ee_2} & \left( 1 + \frac{2d\tau^*}{\tau} p_{-1}\right)\nabla_{1} \mathcal{F}_{\ee_2} & \left( 1 + \frac{2d\tau^*}{\tau} p_{2}\right) (\nabla_{2} +\nabla_{-2}) \mathcal{F}_{\ee_2} -1
\end{pmatrix}.
\end{equation}
 In order to calculate the functions $u_{\rr}$, we start from Eq. (7), and use the following expansions at leading order in $\rho$:
\begin{equation}
\label{ }
\mathcal{A} \sim 2d \left(1+\frac{\tau^*}{\tau}\right) \quad;\quad {A}_\nu \sim 1+\frac{2d\tau^*}{\tau}p_\nu   \quad;\quad   h_\mu \sim v_\mu \rho.
\end{equation}
We then find, at leading order in $\rho$, the equation reduces to
\begin{align}
\label{ }
u_{\rr} = &\frac{1}{2d\left(1+\frac{\tau^*}{\tau}\right)} \left\{  \sum_{\mu} \left(1+\frac{2 d \tau^*}{\tau}p_\mu\right) u_\mu \nabla_{-\mu} \mathcal{F}_{\rr} - \frac{2d\tau^*}{\tau} \sigma [p_1(\nabla_1+1+v_1)-p_{-1}(\nabla_{-1}+1+v_{-1})]  \mathcal{F}_{\rr} \right\} \nonumber\\
&+ \frac{\sigma \frac{\tau^*}{\tau}}{2d \left(1+\frac{\tau^*}{\tau}\right)^2} \left\{\sum_\mu  \left(1+\frac{2 d \tau^*}{\tau}p_\mu\right) v_\mu \nabla_{-\mu} - \frac{2d\tau^*}{\tau} (p_1-p_{-1}) (\nabla_1-\nabla_{-1})  \right\} (p_1\nabla_1-p_{-1}\nabla_{-1}) \mathcal{G}_{\rr}.
\end{align}
Bringing together the terms involving $u_{\rr}$ on the one hand and  $v_{\rr}$ on the other hand, we evaluate this equation for $\rr=\ee_1$, $\ee_{-1}$ and $\ee_2$ and obtain a closed set of three equations that we recast under the matrix form
\begin{equation}
\label{ }
\widetilde{A}\widetilde{u}=\widetilde{C}-\widetilde{B}\widetilde{v}
\end{equation}
with $\widetilde{u}= \begin{pmatrix}
u_1   \\
u_{-1} \\
u_2
\end{pmatrix}$ and
\begin{equation}
\widetilde{A}= \frac{1}{2d\left(1+\frac{\tau^*}{\tau}\right)} \begin{pmatrix}
\left(1+\frac{2d\tau^*}{\tau}p_{1}\right) \nabla_{-1} \mathcal{F}_{\ee_1} -1& \left(1+\frac{2d\tau^*}{\tau}p_{-1}\right) \nabla_{1} \mathcal{F}_{\ee_1} & 2 \left(1+\frac{2d\tau^*}{\tau}p_{2}\right) \nabla_{2} \mathcal{F}_{\ee_1} \\
\left(1+\frac{2d\tau^*}{\tau}p_{1}\right) \nabla_{-1} \mathcal{F}_{\ee_{-1}}  & \left(1+\frac{2d\tau^*}{\tau}p_{-1}\right) \nabla_{1} \mathcal{F}_{\ee_{-1}}-1 & 2\left(1+\frac{2d\tau^*}{\tau}p_{2}\right) \nabla_{2} \mathcal{F}_{\ee_{-1}} \\
\left(1+\frac{2d\tau^*}{\tau}p_{1}\right) \nabla_{-1} \mathcal{F}_{\ee_2} & \left(1+\frac{2d\tau^*}{\tau}p_{-1}\right)\nabla_{1} \mathcal{F}_{\ee_2} &\left(1+\frac{2d\tau^*}{\tau}p_{2}\right) (\nabla_{2} +\nabla_{-2}) \mathcal{F}_{\ee_2} -1
\end{pmatrix},
\end{equation}
{\tiny
\begin{equation}
\widetilde{B}= \frac{\sigma \frac{\tau^*}{\tau}}{1+\frac{\tau^*}{\tau}} 
\begin{pmatrix}
\frac{1+\frac{2d\tau^*}{\tau}p_{1}}{2d\left(  1+ \frac{\tau^*}{\tau}  \right)} \nabla_{-1} (p_1\nabla_1-p_{-1}\nabla_{-1}) \mathcal{G}_{\ee_1} -p_1 \mathcal{F}_{\ee_1}
& \frac{1+\frac{2d\tau^*}{\tau}p_{-1}}{2d\left(  1+ \frac{\tau^*}{\tau}  \right)} \nabla_{1} (p_1\nabla_1-p_{-1}\nabla_{-1}) \mathcal{G}_{\ee_1} +p_{-1} \mathcal{F}_{\ee_1} 
&2\frac{1+\frac{2d\tau^*}{\tau}p_{2}}{2d\left(  1+ \frac{\tau^*}{\tau}  \right)} \nabla_{2} (p_1\nabla_1-p_{-1}\nabla_{-1}) \mathcal{G}_{\ee_1}  \\
\frac{1+\frac{2d\tau^*}{\tau}p_{1}}{2d\left(  1+ \frac{\tau^*}{\tau}  \right)} \nabla_{-1} (p_1\nabla_1-p_{-1}\nabla_{-1}) \mathcal{G}_{\ee_{-1}} -p_1 \mathcal{F}_{\ee_{-1}}
& \frac{1+\frac{2d\tau^*}{\tau}p_{-1}}{2d\left(  1+ \frac{\tau^*}{\tau}  \right)} \nabla_{1} (p_1\nabla_1-p_{-1}\nabla_{-1}) \mathcal{G}_{\ee_{-1}} +p_{-1} \mathcal{F}_{\ee_{-1}} 
&2\frac{1+\frac{2d\tau^*}{\tau}p_{2}}{2d\left(  1+ \frac{\tau^*}{\tau}  \right)} \nabla_{2} (p_1\nabla_1-p_{-1}\nabla_{-1}) \mathcal{G}_{\ee_{-1}}  \\
\frac{1+\frac{2d\tau^*}{\tau}p_{1}}{2d\left(  1+ \frac{\tau^*}{\tau}  \right)} \nabla_{-1} (p_1\nabla_1-p_{-1}\nabla_{-1}) \mathcal{G}_{\ee_2} -p_1 \mathcal{F}_{\ee_2}
& \frac{1+\frac{2d\tau^*}{\tau}p_{-1}}{2d\left(  1+ \frac{\tau^*}{\tau}  \right)} \nabla_{1} (p_1\nabla_1-p_{-1}\nabla_{-1}) \mathcal{G}_{\ee_2} +p_{-1} \mathcal{F}_{\ee_2} 
&\frac{1+\frac{2d\tau^*}{\tau}p_{2}}{2d\left(  1+ \frac{\tau^*}{\tau}  \right)} (\nabla_{2}+\nabla_{-2}) (p_1\nabla_1-p_{-1}\nabla_{-1}) \mathcal{G}_{\ee_2}  \\
\end{pmatrix},
\end{equation}
}
\begin{equation}
\widetilde{C}= 
\begin{pmatrix}
\frac{\frac{\tau^*}{\tau}}{1+ \frac{\tau^*}{\tau}  } \sigma [p_1 (\nabla_1+1)-p_{-1}(\nabla_{-1}+1)] \mathcal{F}_{\ee_1}   +\left( \frac{\frac{\tau^*}{\tau}}{1+ \frac{\tau^*}{\tau}  }\right)^2 \sigma (p_1-p_{-1}) (\nabla_1-\nabla_{-1}) (p_1 \nabla_1 -p_{-1} \nabla_{-1})  \mathcal{G}_{\ee_1} \\
\frac{\frac{\tau^*}{\tau}}{1+ \frac{\tau^*}{\tau}  } \sigma [p_1 (\nabla_1+1)-p_{-1}(\nabla_{-1}+1)] \mathcal{F}_{\ee_{-1}}   +\left( \frac{\frac{\tau^*}{\tau}}{1+ \frac{\tau^*}{\tau}  }\right)^2 \sigma (p_1-p_{-1}) (\nabla_1-\nabla_{-1}) (p_1 \nabla_1 -p_{-1} \nabla_{-1})  \mathcal{G}_{\ee_{-1}}\\
\frac{\frac{\tau^*}{\tau}}{1+ \frac{\tau^*}{\tau}  } \sigma [p_1 (\nabla_1+1)-p_{-1}(\nabla_{-1}+1)] \mathcal{F}_{\ee_2}   +\left( \frac{\frac{\tau^*}{\tau}}{1+ \frac{\tau^*}{\tau}  }\right)^2 \sigma (p_1-p_{-1}) (\nabla_1-\nabla_{-1}) (p_1 \nabla_1 -p_{-1} \nabla_{-1})  \mathcal{G}_{\ee_2}\\
\end{pmatrix}.
\end{equation}
It is easy to take the limit of fixed obstacles in these equations ($\tau^*\to\infty$) to retrieve the results concerning the Lorentz gas. In this particular limit, we notice $\lim_{\tau^*\to\infty} \widetilde{A}=\widetilde{D}$.  Finally, $u_1$ and $u_{-1}$ are obtained with
\begin{equation}
\label{umatrix}
\begin{pmatrix}
u_1   \\
u_{-1} \\
u_2
\end{pmatrix} = \widetilde{A}^{-1}(\widetilde{C}-\widetilde{B}\widetilde{v}),
\end{equation}
and are used to calculate the diffusion coefficient using Eq. \eqref{K_small_rho_exp}.


\begin{thebibliography}{36}
\expandafter\ifx\csname natexlab\endcsname\relax\def\natexlab#1{#1}\fi
\expandafter\ifx\csname bibnamefont\endcsname\relax
  \def\bibnamefont#1{#1}\fi
\expandafter\ifx\csname bibfnamefont\endcsname\relax
  \def\bibfnamefont#1{#1}\fi
\expandafter\ifx\csname citenamefont\endcsname\relax
  \def\citenamefont#1{#1}\fi
\expandafter\ifx\csname url\endcsname\relax
  \def\url#1{\texttt{#1}}\fi
\expandafter\ifx\csname urlprefix\endcsname\relax\def\urlprefix{URL }\fi
\providecommand{\bibinfo}[2]{#2}
\providecommand{\eprint}[2][]{\url{#2}}

\bibitem[{\citenamefont{H{\"{o}}fling and Franosch}(2013)}]{Hofling2013}
\bibinfo{author}{\bibfnamefont{F.}~\bibnamefont{H{\"{o}}fling}}
  \bibnamefont{and} \bibinfo{author}{\bibfnamefont{T.}~\bibnamefont{Franosch}},
  \bibinfo{journal}{Rep. Prog. Phys.} \textbf{\bibinfo{volume}{76}},
  \bibinfo{pages}{046602} (\bibinfo{year}{2013}).

\bibitem[{\citenamefont{Chou et~al.}(2011)\citenamefont{Chou, Mallick, and
  Zia}}]{Chou2011a}
\bibinfo{author}{\bibfnamefont{T.}~\bibnamefont{Chou}},
  \bibinfo{author}{\bibfnamefont{K.}~\bibnamefont{Mallick}}, \bibnamefont{and}
  \bibinfo{author}{\bibfnamefont{R.~K.~P.} \bibnamefont{Zia}},
  \bibinfo{journal}{Rep. Prog. Phys.} \textbf{\bibinfo{volume}{74}},
  \bibinfo{pages}{116601} (\bibinfo{year}{2011}).

\bibitem[{\citenamefont{Bechinger et~al.}(2016)\citenamefont{Bechinger, {Di
  Leonardo}, L{\"{o}}wen, Reichhardt, Volpe, and Volpe}}]{Bechinger2016}
\bibinfo{author}{\bibfnamefont{C.}~\bibnamefont{Bechinger}},
  \bibinfo{author}{\bibfnamefont{R.}~\bibnamefont{{Di Leonardo}}},
  \bibinfo{author}{\bibfnamefont{H.}~\bibnamefont{L{\"{o}}wen}},
  \bibinfo{author}{\bibfnamefont{C.}~\bibnamefont{Reichhardt}},
  \bibinfo{author}{\bibfnamefont{G.}~\bibnamefont{Volpe}}, \bibnamefont{and}
  \bibinfo{author}{\bibfnamefont{G.}~\bibnamefont{Volpe}},
  \bibinfo{journal}{Rev. Mod. Phys.} \textbf{\bibinfo{volume}{88}},
  \bibinfo{pages}{045006} (\bibinfo{year}{2016}).

\bibitem[{\citenamefont{Illien et~al.}(2017)\citenamefont{Illien, Golestanian,
  and Sen}}]{Illien2017}
\bibinfo{author}{\bibfnamefont{P.}~\bibnamefont{Illien}},
  \bibinfo{author}{\bibfnamefont{R.}~\bibnamefont{Golestanian}},
  \bibnamefont{and} \bibinfo{author}{\bibfnamefont{A.}~\bibnamefont{Sen}},
  \bibinfo{journal}{Chem. Soc. Rev.} \textbf{\bibinfo{volume}{46}},
  \bibinfo{pages}{5508} (\bibinfo{year}{2017}).

\bibitem[{\citenamefont{Wilson and Poon}(2011)}]{Wilson2011}
\bibinfo{author}{\bibfnamefont{L.~G.} \bibnamefont{Wilson}} \bibnamefont{and}
  \bibinfo{author}{\bibfnamefont{W.~C.~K.} \bibnamefont{Poon}},
  \bibinfo{journal}{Phys. Chem. Chem. Phys.} \textbf{\bibinfo{volume}{13}},
  \bibinfo{pages}{10617} (\bibinfo{year}{2011}).

\bibitem[{\citenamefont{Dullens and Bechinger}(2011)}]{Dullens2011}
\bibinfo{author}{\bibfnamefont{R.~P.~A.} \bibnamefont{Dullens}}
  \bibnamefont{and}
  \bibinfo{author}{\bibfnamefont{C.}~\bibnamefont{Bechinger}},
  \bibinfo{journal}{Phys. Rev. Lett.} \textbf{\bibinfo{volume}{107}},
  \bibinfo{pages}{138301} (\bibinfo{year}{2011}).

\bibitem[{\citenamefont{Puertas and Voigtmann}(2014)}]{Puertas2014a}
\bibinfo{author}{\bibfnamefont{A.~M.} \bibnamefont{Puertas}} \bibnamefont{and}
  \bibinfo{author}{\bibfnamefont{T.}~\bibnamefont{Voigtmann}},
  \bibinfo{journal}{J. Phys. Condens. Matter} \textbf{\bibinfo{volume}{26}},
  \bibinfo{pages}{243101} (\bibinfo{year}{2014}).

\bibitem[{\citenamefont{Mejia-Monasterio and
  Oshanin}(2010)}]{Mejia-Monasterio2010}
\bibinfo{author}{\bibfnamefont{C.}~\bibnamefont{Mejia-Monasterio}}
  \bibnamefont{and} \bibinfo{author}{\bibfnamefont{G.}~\bibnamefont{Oshanin}},
  \bibinfo{journal}{Soft Matter} \textbf{\bibinfo{volume}{7}},
  \bibinfo{pages}{993} (\bibinfo{year}{2010}).

\bibitem[{\citenamefont{Tanaka et~al.}(2017)\citenamefont{Tanaka, Lee, and
  Brenner}}]{Tanaka2016a}
\bibinfo{author}{\bibfnamefont{H.}~\bibnamefont{Tanaka}},
  \bibinfo{author}{\bibfnamefont{A.~A.} \bibnamefont{Lee}}, \bibnamefont{and}
  \bibinfo{author}{\bibfnamefont{M.~P.} \bibnamefont{Brenner}},
  \bibinfo{journal}{Phys. Rev. Fluids} \textbf{\bibinfo{volume}{2}},
  \bibinfo{pages}{043103} (\bibinfo{year}{2017}).

\bibitem[{\citenamefont{Weber et~al.}(2016)\citenamefont{Weber, Weber, and
  Frey}}]{Weber2016}
\bibinfo{author}{\bibfnamefont{S.~N.} \bibnamefont{Weber}},
  \bibinfo{author}{\bibfnamefont{C.~A.} \bibnamefont{Weber}}, \bibnamefont{and}
  \bibinfo{author}{\bibfnamefont{E.}~\bibnamefont{Frey}},
  \bibinfo{journal}{Phys. Rev. Lett.} \textbf{\bibinfo{volume}{116}},
  \bibinfo{pages}{058301} (\bibinfo{year}{2016}).

\bibitem[{\citenamefont{Stenhammar et~al.}(2015)\citenamefont{Stenhammar,
  Wittkowski, Marenduzzo, and Cates}}]{Stenhammar2015}
\bibinfo{author}{\bibfnamefont{J.}~\bibnamefont{Stenhammar}},
  \bibinfo{author}{\bibfnamefont{R.}~\bibnamefont{Wittkowski}},
  \bibinfo{author}{\bibfnamefont{D.}~\bibnamefont{Marenduzzo}},
  \bibnamefont{and} \bibinfo{author}{\bibfnamefont{M.~E.} \bibnamefont{Cates}},
  \bibinfo{journal}{Phys. Rev. Lett.} \textbf{\bibinfo{volume}{114}},
  \bibinfo{pages}{018301} (\bibinfo{year}{2015}).

\bibitem[{\citenamefont{D{\'{e}}mery}(2015)}]{Demery2015}
\bibinfo{author}{\bibfnamefont{V.}~\bibnamefont{D{\'{e}}mery}},
  \bibinfo{journal}{Phys. Rev. E} \textbf{\bibinfo{volume}{91}},
  \bibinfo{pages}{062301} (\bibinfo{year}{2015}).

\bibitem[{\citenamefont{D{\'{e}}mery et~al.}(2014)\citenamefont{D{\'{e}}mery,
  B{\'{e}}nichou, and Jacquin}}]{Demery2014}
\bibinfo{author}{\bibfnamefont{V.}~\bibnamefont{D{\'{e}}mery}},
  \bibinfo{author}{\bibfnamefont{O.}~\bibnamefont{B{\'{e}}nichou}},
  \bibnamefont{and} \bibinfo{author}{\bibfnamefont{H.}~\bibnamefont{Jacquin}},
  \bibinfo{journal}{New J. Phys.} \textbf{\bibinfo{volume}{16}},
  \bibinfo{pages}{053032} (\bibinfo{year}{2014}).

\bibitem[{\citenamefont{Mallick}(2015)}]{Mallick2015}
\bibinfo{author}{\bibfnamefont{K.}~\bibnamefont{Mallick}},
  \bibinfo{journal}{Physica A} \textbf{\bibinfo{volume}{418}},
  \bibinfo{pages}{17} (\bibinfo{year}{2015}).

\bibitem[{\citenamefont{B{\'{e}}nichou
  et~al.}(2000)\citenamefont{B{\'{e}}nichou, Cazabat, {De Coninck}, Moreau, and
  Oshanin}}]{Benichou1999a}
\bibinfo{author}{\bibfnamefont{O.}~\bibnamefont{B{\'{e}}nichou}},
  \bibinfo{author}{\bibfnamefont{A.~M.} \bibnamefont{Cazabat}},
  \bibinfo{author}{\bibfnamefont{J.}~\bibnamefont{{De Coninck}}},
  \bibinfo{author}{\bibfnamefont{M.}~\bibnamefont{Moreau}}, \bibnamefont{and}
  \bibinfo{author}{\bibfnamefont{G.}~\bibnamefont{Oshanin}},
  \bibinfo{journal}{Phys. Rev. Lett.} \textbf{\bibinfo{volume}{84}},
  \bibinfo{pages}{511} (\bibinfo{year}{2000}).

\bibitem[{\citenamefont{B{\'{e}}nichou
  et~al.}(2001)\citenamefont{B{\'{e}}nichou, Cazabat, {De Coninck}, Moreau, and
  Oshanin}}]{Benichou2001}
\bibinfo{author}{\bibfnamefont{O.}~\bibnamefont{B{\'{e}}nichou}},
  \bibinfo{author}{\bibfnamefont{A.~M.} \bibnamefont{Cazabat}},
  \bibinfo{author}{\bibfnamefont{J.}~\bibnamefont{{De Coninck}}},
  \bibinfo{author}{\bibfnamefont{M.}~\bibnamefont{Moreau}}, \bibnamefont{and}
  \bibinfo{author}{\bibfnamefont{G.}~\bibnamefont{Oshanin}},
  \bibinfo{journal}{Phys. Rev. B} \textbf{\bibinfo{volume}{63}},
  \bibinfo{pages}{235413} (\bibinfo{year}{2001}).

\bibitem[{\citenamefont{Leitmann and Franosch}(2013)}]{Leitmann2013}
\bibinfo{author}{\bibfnamefont{S.}~\bibnamefont{Leitmann}} \bibnamefont{and}
  \bibinfo{author}{\bibfnamefont{T.}~\bibnamefont{Franosch}},
  \bibinfo{journal}{Phys. Rev. Lett.} \textbf{\bibinfo{volume}{111}},
  \bibinfo{pages}{190603} (\bibinfo{year}{2013}).

\bibitem[{\citenamefont{B{\'{e}}nichou
  et~al.}(2016)\citenamefont{B{\'{e}}nichou, Illien, Oshanin, Sarracino, and
  Voituriez}}]{Benichou2015c}
\bibinfo{author}{\bibfnamefont{O.}~\bibnamefont{B{\'{e}}nichou}},
  \bibinfo{author}{\bibfnamefont{P.}~\bibnamefont{Illien}},
  \bibinfo{author}{\bibfnamefont{G.}~\bibnamefont{Oshanin}},
  \bibinfo{author}{\bibfnamefont{A.}~\bibnamefont{Sarracino}},
  \bibnamefont{and}
  \bibinfo{author}{\bibfnamefont{R.}~\bibnamefont{Voituriez}},
  \bibinfo{journal}{Phys. Rev. E} \textbf{\bibinfo{volume}{93}},
  \bibinfo{pages}{032128} (\bibinfo{year}{2016}).

\bibitem[{\citenamefont{Leitmann and Franosch}(2017)}]{Leitmann2017}
\bibinfo{author}{\bibfnamefont{S.}~\bibnamefont{Leitmann}} \bibnamefont{and}
  \bibinfo{author}{\bibfnamefont{T.}~\bibnamefont{Franosch}},
  \bibinfo{journal}{Phys. Rev. Lett.} \textbf{\bibinfo{volume}{118}},
  \bibinfo{pages}{018001} (\bibinfo{year}{2017}).

\bibitem[{\citenamefont{B{\'{e}}nichou
  et~al.}(2013)\citenamefont{B{\'{e}}nichou, Bodrova, Chakraborty, Illien, Law,
  Mej{\'{i}}a-Monasterio, Oshanin, and Voituriez}}]{Benichou2013c}
\bibinfo{author}{\bibfnamefont{O.}~\bibnamefont{B{\'{e}}nichou}},
  \bibinfo{author}{\bibfnamefont{A.}~\bibnamefont{Bodrova}},
  \bibinfo{author}{\bibfnamefont{D.}~\bibnamefont{Chakraborty}},
  \bibinfo{author}{\bibfnamefont{P.}~\bibnamefont{Illien}},
  \bibinfo{author}{\bibfnamefont{A.}~\bibnamefont{Law}},
  \bibinfo{author}{\bibfnamefont{C.}~\bibnamefont{Mej{\'{i}}a-Monasterio}},
  \bibinfo{author}{\bibfnamefont{G.}~\bibnamefont{Oshanin}}, \bibnamefont{and}
  \bibinfo{author}{\bibfnamefont{R.}~\bibnamefont{Voituriez}},
  \bibinfo{journal}{Phys. Rev. Lett.} \textbf{\bibinfo{volume}{111}},
  \bibinfo{pages}{260601} (\bibinfo{year}{2013}).

\bibitem[{\citenamefont{Winter et~al.}(2012)\citenamefont{Winter, Horbach,
  Virnau, and Binder}}]{Winter2012}
\bibinfo{author}{\bibfnamefont{D.}~\bibnamefont{Winter}},
  \bibinfo{author}{\bibfnamefont{J.}~\bibnamefont{Horbach}},
  \bibinfo{author}{\bibfnamefont{P.}~\bibnamefont{Virnau}}, \bibnamefont{and}
  \bibinfo{author}{\bibfnamefont{K.}~\bibnamefont{Binder}},
  \bibinfo{journal}{Phys. Rev. Lett.} \textbf{\bibinfo{volume}{108}},
  \bibinfo{pages}{028303} (\bibinfo{year}{2012}).

\bibitem[{\citenamefont{Schroer and Heuer}(2013)}]{Schroer2013}
\bibinfo{author}{\bibfnamefont{C.~F.~E.} \bibnamefont{Schroer}}
  \bibnamefont{and} \bibinfo{author}{\bibfnamefont{A.}~\bibnamefont{Heuer}},
  \bibinfo{journal}{Phys. Rev. Lett.} \textbf{\bibinfo{volume}{110}},
  \bibinfo{pages}{067801} (\bibinfo{year}{2013}).

\bibitem[{\citenamefont{Reimann et~al.}(2001)\citenamefont{Reimann, {Van den
  Broeck}, Linke, H{\"{a}}nggi, Rubi, and P{\'{e}}rez-Madrid}}]{Reimann2001}
\bibinfo{author}{\bibfnamefont{P.}~\bibnamefont{Reimann}},
  \bibinfo{author}{\bibfnamefont{C.}~\bibnamefont{{Van den Broeck}}},
  \bibinfo{author}{\bibfnamefont{H.}~\bibnamefont{Linke}},
  \bibinfo{author}{\bibfnamefont{P.}~\bibnamefont{H{\"{a}}nggi}},
  \bibinfo{author}{\bibfnamefont{J.~M.} \bibnamefont{Rubi}}, \bibnamefont{and}
  \bibinfo{author}{\bibfnamefont{A.}~\bibnamefont{P{\'{e}}rez-Madrid}},
  \bibinfo{journal}{Phys. Rev. Lett.} \textbf{\bibinfo{volume}{87}},
  \bibinfo{pages}{010602} (\bibinfo{year}{2001}).

\bibitem[{\citenamefont{Lindenberg et~al.}(2005)\citenamefont{Lindenberg,
  Lacasta, Sancho, and Romero}}]{Lindenberg2005}
\bibinfo{author}{\bibfnamefont{K.}~\bibnamefont{Lindenberg}},
  \bibinfo{author}{\bibfnamefont{A.~M.} \bibnamefont{Lacasta}},
  \bibinfo{author}{\bibfnamefont{J.~M.} \bibnamefont{Sancho}},
  \bibnamefont{and} \bibinfo{author}{\bibfnamefont{A.~H.}
  \bibnamefont{Romero}}, \bibinfo{journal}{New J. Phys.}
  \textbf{\bibinfo{volume}{7}}, \bibinfo{pages}{29} (\bibinfo{year}{2005}).

\bibitem[{\citenamefont{Lindner and Sokolov}(2016)}]{Lindner2016}
\bibinfo{author}{\bibfnamefont{B.}~\bibnamefont{Lindner}} \bibnamefont{and}
  \bibinfo{author}{\bibfnamefont{I.~M.} \bibnamefont{Sokolov}},
  \bibinfo{journal}{Phys. Rev. E} \textbf{\bibinfo{volume}{93}},
  \bibinfo{pages}{042106} (\bibinfo{year}{2016}).

\bibitem[{\citenamefont{Reimann and Eichhorn}(2008)}]{Reimann2008}
\bibinfo{author}{\bibfnamefont{P.}~\bibnamefont{Reimann}} \bibnamefont{and}
  \bibinfo{author}{\bibfnamefont{R.}~\bibnamefont{Eichhorn}},
  \bibinfo{journal}{Phys. Rev. Lett.} \textbf{\bibinfo{volume}{101}},
  \bibinfo{pages}{180601} (\bibinfo{year}{2008}).

\bibitem[{\citenamefont{Lindner and Nicola}(2008)}]{Lindner2008}
\bibinfo{author}{\bibfnamefont{B.}~\bibnamefont{Lindner}} \bibnamefont{and}
  \bibinfo{author}{\bibfnamefont{E.~M.} \bibnamefont{Nicola}},
  \bibinfo{journal}{Phys. Rev. Lett.} \textbf{\bibinfo{volume}{101}},
  \bibinfo{pages}{190603} (\bibinfo{year}{2008}).

\bibitem[{\citenamefont{Nakazato and Kitahara}(1980)}]{Nakazato1980}
\bibinfo{author}{\bibfnamefont{K.}~\bibnamefont{Nakazato}} \bibnamefont{and}
  \bibinfo{author}{\bibfnamefont{K.}~\bibnamefont{Kitahara}},
  \bibinfo{journal}{Prog. Theor. Phys.} \textbf{\bibinfo{volume}{64}},
  \bibinfo{pages}{2261} (\bibinfo{year}{1980}).

\bibitem[{\citenamefont{{Supplementary Information}}()}]{SI}
\bibinfo{author}{\bibnamefont{{Supplemental Material}}}

\bibitem[{\citenamefont{B{\'{e}}nichou
  et~al.}(2014)\citenamefont{B{\'{e}}nichou, Illien, Oshanin, Sarracino, and
  Voituriez}}]{Benichou2014}
\bibinfo{author}{\bibfnamefont{O.}~\bibnamefont{B{\'{e}}nichou}},
  \bibinfo{author}{\bibfnamefont{P.}~\bibnamefont{Illien}},
  \bibinfo{author}{\bibfnamefont{G.}~\bibnamefont{Oshanin}},
  \bibinfo{author}{\bibfnamefont{A.}~\bibnamefont{Sarracino}},
  \bibnamefont{and}
  \bibinfo{author}{\bibfnamefont{R.}~\bibnamefont{Voituriez}},
  \bibinfo{journal}{Phys. Rev. Lett.} \textbf{\bibinfo{volume}{113}},
  \bibinfo{pages}{268002} (\bibinfo{year}{2014}).

\bibitem[{\citenamefont{Illien et~al.}(2015)\citenamefont{Illien,
  B{\'{e}}nichou, Oshanin, and Voituriez}}]{Illien2015}
\bibinfo{author}{\bibfnamefont{P.}~\bibnamefont{Illien}},
  \bibinfo{author}{\bibfnamefont{O.}~\bibnamefont{B{\'{e}}nichou}},
  \bibinfo{author}{\bibfnamefont{G.}~\bibnamefont{Oshanin}}, \bibnamefont{and}
  \bibinfo{author}{\bibfnamefont{R.}~\bibnamefont{Voituriez}},
  \bibinfo{journal}{J. Stat. Mech.} p. \bibinfo{pages}{P11016}
  (\bibinfo{year}{2015}).

\bibitem[{\citenamefont{Hughes}(1995)}]{Hughes1995}
\bibinfo{author}{\bibfnamefont{B.~D.} \bibnamefont{Hughes}},
  \emph{\bibinfo{title}{{Random Walks and Random Environments: Random walks,
  Volume 1}}} (\bibinfo{publisher}{Oxford University Press},
  \bibinfo{address}{Oxford}, \bibinfo{year}{1995}).


  \bibitem{note} We checked numerically that $D(F)$ is decreasing even for $\rho=0.1$ and $\tau^*=10$, but too slowly to be shown with this scale.



\bibitem[{\citenamefont{Baerts et~al.}(2013)\citenamefont{Baerts, Basu, Maes,
  and Safaverdi}}]{Baerts2013}
\bibinfo{author}{\bibfnamefont{P.}~\bibnamefont{Baerts}},
  \bibinfo{author}{\bibfnamefont{U.}~\bibnamefont{Basu}},
  \bibinfo{author}{\bibfnamefont{C.}~\bibnamefont{Maes}}, \bibnamefont{and}
  \bibinfo{author}{\bibfnamefont{S.}~\bibnamefont{Safaverdi}},
  \bibinfo{journal}{Phys. Rev. E} \textbf{\bibinfo{volume}{88}},
  \bibinfo{pages}{052109} (\bibinfo{year}{2013}).

\bibitem[{\citenamefont{Brummelhuis and Hilhorst}(1988)}]{Brummelhuis1988}
\bibinfo{author}{\bibfnamefont{M.~J. A.~M.} \bibnamefont{Brummelhuis}}
  \bibnamefont{and} \bibinfo{author}{\bibfnamefont{H.~J.}
  \bibnamefont{Hilhorst}}, \bibinfo{journal}{J. Stat. Phys.}
  \textbf{\bibinfo{volume}{53}}, \bibinfo{pages}{249} (\bibinfo{year}{1988}).

\bibitem[{\citenamefont{Brummelhuis and Hilhorst}(1989)}]{Brummelhuis1989a}
\bibinfo{author}{\bibfnamefont{M.~J. A.~M.} \bibnamefont{Brummelhuis}}
  \bibnamefont{and} \bibinfo{author}{\bibfnamefont{H.~J.}
  \bibnamefont{Hilhorst}}, \bibinfo{journal}{Physica A}
  \textbf{\bibinfo{volume}{156}}, \bibinfo{pages}{575} (\bibinfo{year}{1989}).



\bibitem[{\citenamefont{B{\'{e}}nichou and Oshanin}(2002)}]{Benichou2002a}
\bibinfo{author}{\bibfnamefont{O.}~\bibnamefont{B{\'{e}}nichou}}
  \bibnamefont{and} \bibinfo{author}{\bibfnamefont{G.}~\bibnamefont{Oshanin}},
  \bibinfo{journal}{Phys. Rev. E} \textbf{\bibinfo{volume}{66}},
  \bibinfo{pages}{031101} (\bibinfo{year}{2002}).
  

  
  

\end{thebibliography}

\begin{thebibliography}{5}%
\makeatletter
\providecommand \@ifxundefined [1]{%
 \@ifx{#1\undefined}
}%
\providecommand \@ifnum [1]{%
 \ifnum #1\expandafter \@firstoftwo
 \else \expandafter \@secondoftwo
 \fi
}%
\providecommand \@ifx [1]{%
 \ifx #1\expandafter \@firstoftwo
 \else \expandafter \@secondoftwo
 \fi
}%
\providecommand \natexlab [1]{#1}%
\providecommand \enquote  [1]{``#1''}%
\providecommand \bibnamefont  [1]{#1}%
\providecommand \bibfnamefont [1]{#1}%
\providecommand \citenamefont [1]{#1}%
\providecommand \href@noop [0]{\@secondoftwo}%
\providecommand \href [0]{\begingroup \@sanitize@url \@href}%
\providecommand \@href[1]{\@@startlink{#1}\@@href}%
\providecommand \@@href[1]{\endgroup#1\@@endlink}%
\providecommand \@sanitize@url [0]{\catcode `\\12\catcode `\$12\catcode
  `\&12\catcode `\#12\catcode `\^12\catcode `\_12\catcode `\%12\relax}%
\providecommand \@@startlink[1]{}%
\providecommand \@@endlink[0]{}%
\providecommand \url  [0]{\begingroup\@sanitize@url \@url }%
\providecommand \@url [1]{\endgroup\@href {#1}{\urlprefix }}%
\providecommand \urlprefix  [0]{URL }%
\providecommand \Eprint [0]{\href }%
\providecommand \doibase [0]{http://dx.doi.org/}%
\providecommand \selectlanguage [0]{\@gobble}%
\providecommand \bibinfo  [0]{\@secondoftwo}%
\providecommand \bibfield  [0]{\@secondoftwo}%
\providecommand \translation [1]{[#1]}%
\providecommand \BibitemOpen [0]{}%
\providecommand \bibitemStop [0]{}%
\providecommand \bibitemNoStop [0]{.\EOS\space}%
\providecommand \EOS [0]{\spacefactor3000\relax}%
\providecommand \BibitemShut  [1]{\csname bibitem#1\endcsname}%
\let\auto@bib@innerbib\@empty
\bibitem [{\citenamefont {Illien}\ \emph {et~al.}(2015)\citenamefont {Illien},
  \citenamefont {B{\'{e}}nichou}, \citenamefont {Oshanin},\ and\ \citenamefont
  {Voituriez}}]{Illien2015}%
  \BibitemOpen
  \bibfield  {author} {\bibinfo {author} {\bibfnamefont {P.}~\bibnamefont
  {Illien}}, \bibinfo {author} {\bibfnamefont {O.}~\bibnamefont
  {B{\'{e}}nichou}}, \bibinfo {author} {\bibfnamefont {G.}~\bibnamefont
  {Oshanin}}, \ and\ \bibinfo {author} {\bibfnamefont {R.}~\bibnamefont
  {Voituriez}},\ }\href {\doibase 10.1088/1742-5468/15/11/P11016} {\bibfield
  {journal} {\bibinfo  {journal} {J. Stat. Mech.} \bibinfo {pages}
  {P11016}} (\bibinfo {year} {2015})}\BibitemShut {NoStop}%
\bibitem [{\citenamefont {Hughes}(1995)}]{Hughes1995}%
  \BibitemOpen
  \bibfield  {author} {\bibinfo {author} {\bibfnamefont {B.~D.}\ \bibnamefont
  {Hughes}},\ }\href@noop {} {\emph {\bibinfo {title} {{Random Walks and Random
  Environments: Random walks, Volume 1}}}}\ (\bibinfo  {publisher} {Oxford
  University Press},\ \bibinfo {address} {Oxford},\ \bibinfo {year}
  {1995})\BibitemShut {NoStop}%
\bibitem [{\citenamefont {Brummelhuis}\ and\ \citenamefont
  {Hilhorst}(1988)}]{Brummelhuis1988}%
  \BibitemOpen
  \bibfield  {author} {\bibinfo {author} {\bibfnamefont {M.~J. A.~M.}\
  \bibnamefont {Brummelhuis}}\ and\ \bibinfo {author} {\bibfnamefont {H.~J.}\
  \bibnamefont {Hilhorst}},\ }\href {\doibase 10.1007/BF01011556} {\bibfield
  {journal} {\bibinfo  {journal} {J. Stat. Phys.}\ }\textbf {\bibinfo {volume}
  {53}},\ \bibinfo {pages} {249} (\bibinfo {year} {1988})}\BibitemShut
  {NoStop}%
\bibitem [{\citenamefont {Illien}\ \emph {et~al.}(2014)\citenamefont {Illien},
  \citenamefont {B{\'{e}}nichou}, \citenamefont {Oshanin},\ and\ \citenamefont
  {Voituriez}}]{Illien2014}%
  \BibitemOpen
  \bibfield  {author} {\bibinfo {author} {\bibfnamefont {P.}~\bibnamefont
  {Illien}}, \bibinfo {author} {\bibfnamefont {O.}~\bibnamefont
  {B{\'{e}}nichou}}, \bibinfo {author} {\bibfnamefont {G.}~\bibnamefont
  {Oshanin}}, \ and\ \bibinfo {author} {\bibfnamefont {R.}~\bibnamefont
  {Voituriez}},\ }\href {\doibase 10.1103/PhysRevLett.113.030603} {\bibfield
  {journal} {\bibinfo  {journal} {Phys. Rev. Lett.}\ }\textbf {\bibinfo
  {volume} {113}},\ \bibinfo {pages} {030603} (\bibinfo {year}
  {2014})}\BibitemShut {NoStop}%
\bibitem [{\citenamefont {B{\'{e}}nichou}\ \emph {et~al.}(2014)\citenamefont
  {B{\'{e}}nichou}, \citenamefont {Illien}, \citenamefont {Oshanin},
  \citenamefont {Sarracino},\ and\ \citenamefont {Voituriez}}]{Benichou2014}%
  \BibitemOpen
  \bibfield  {author} {\bibinfo {author} {\bibfnamefont {O.}~\bibnamefont
  {B{\'{e}}nichou}}, \bibinfo {author} {\bibfnamefont {P.}~\bibnamefont
  {Illien}}, \bibinfo {author} {\bibfnamefont {G.}~\bibnamefont {Oshanin}},
  \bibinfo {author} {\bibfnamefont {A.}~\bibnamefont {Sarracino}}, \ and\
  \bibinfo {author} {\bibfnamefont {R.}~\bibnamefont {Voituriez}},\ }\href
  {\doibase 10.1103/PhysRevLett.113.268002} {\bibfield  {journal} {\bibinfo
  {journal} {Phys. Rev. Lett.}\ }\textbf {\bibinfo {volume} {113}},\ \bibinfo
  {pages} {268002} (\bibinfo {year} {2014})}\BibitemShut {NoStop}%
\end{thebibliography}

%

\end{document}